\def\standardrisposta{s }\def\reducedrisposta{r }
\def\mplarisposta{mpla }\def\zerorisposta{z }\def\bigrisposta{big }
\def\doublerisposta{d }\def\cartarisposta{e }\def\amsrisposta{y }
\newcount\ingrandimento \newcount\sinnota \newcount\dimnota
\newcount\unoduecol \newdimen\collhsize \newdimen\tothsize
\newdimen\fullhsize \newcount\controllorisposta \sinnota=1
\newskip\infralinea  \global\controllorisposta=0
\immediate\write16 { ********  Welcome to PANDA macros (Plain TeX,
AP, 1991) ******** }
%
%
%
%
\immediate\write16 { You'll have to answer a few questions in
lowercase.}
\message{>  Do you want it in double-page (d), reduced (r)
or standard format (s) ? }\read-1 to\risposta
\message{>  Do you want it in USA A4 (u) or EUROPEAN A4
(e) paper size ? }\read-1 to\srisposta
\message{>  Do you have AMSFonts 2.0 (math) fonts (y/n) ? }
\read-1 to\arisposta
%
%
%
%
%
%
%
%
%
\def\srisposta{e } \def\arisposta{y }
\ifx\risposta\standardrisposta \ingrandimento=1200
\message {>> This will come out UNREDUCED << }
\dimnota=2 \unoduecol=1 \global\controllorisposta=1 \fi
\ifx\risposta\bigrisposta \ingrandimento=1440
\message {>> This will come out ENLARGED << }
\dimnota=2 \unoduecol=1 \global\controllorisposta=1 \fi
\ifx\risposta\reducedrisposta \ingrandimento=1095 \dimnota=1
\unoduecol=1  \global\controllorisposta=1
\message {>> This will come out REDUCED << } \fi
\ifx\risposta\doublerisposta \ingrandimento=1000 \dimnota=2
\unoduecol=2  \message {>> You must print this in
LANDSCAPE orientation << } \global\controllorisposta=1 \fi
\ifx\risposta\mplarisposta \ingrandimento=1000 \dimnota=1
\message {>> Mod. Phys. Lett. A format << }
\unoduecol=1 \global\controllorisposta=1 \fi
\ifx\risposta\zerorisposta \ingrandimento=1000 \dimnota=2
\message {>> Zero Magnification format << }
\unoduecol=1 \global\controllorisposta=1 \fi
\ifnum\controllorisposta=0  \ingrandimento=1200
\message {>>> ERROR IN INPUT, I ASSUME STANDARD
UNREDUCED FORMAT <<< }  \dimnota=2 \unoduecol=1 \fi
\magnification=\ingrandimento
%
%
%
%
\newdimen\eucolumnsize \newdimen\eudoublehsize \newdimen\eudoublevsize
\newdimen\uscolumnsize \newdimen\usdoublehsize \newdimen\usdoublevsize
\newdimen\eusinglehsize \newdimen\eusinglevsize \newdimen\ussinglehsize
\newskip\standardbaselineskip \newdimen\ussinglevsize
\newskip\reducedbaselineskip \newskip\doublebaselineskip
\newskip\bigbaselineskip
\eucolumnsize=12.0truecm    
\eudoublehsize=25.5truecm   
\eudoublevsize=6.5truein    
\uscolumnsize=4.4truein     
\usdoublehsize=9.4truein    
\usdoublevsize=6.8truein    
\eusinglehsize=6.3truein    
\eusinglevsize=24truecm     
\ussinglehsize=6.5truein    
\ussinglevsize=8.9truein    
\bigbaselineskip=18pt plus.2pt       
\standardbaselineskip=16pt plus.2pt  
\reducedbaselineskip=14pt plus.2pt   
\doublebaselineskip=12pt plus.2pt    
%
%
\def\Portoffset{}
\def\Landoffset{\hoffset=-.140truein}
\ifx\risposta\mplarisposta \def\Portoffset{\hoffset=1.9truecm
\voffset=1.4truecm} \fi
%
%
\def\Landspec{}
\tolerance=10000
\parskip=0pt plus2pt  \leftskip=0pt \rightskip=0pt
%
%
\ifx\risposta\bigrisposta      \infralinea=\bigbaselineskip \fi
\ifx\risposta\standardrisposta \infralinea=\standardbaselineskip \fi
\ifx\risposta\reducedrisposta  \infralinea=\reducedbaselineskip \fi
\ifx\risposta\doublerisposta   \infralinea=\doublebaselineskip \fi
\ifx\risposta\mplarisposta     \infralinea=13pt \fi
\ifx\risposta\zerorisposta     \infralinea=12pt plus.2pt\fi
\ifnum\controllorisposta=0    \infralinea=\standardbaselineskip \fi
\ifx\risposta\doublerisposta   \Landoffset \else \Portoffset \fi
\ifx\risposta\doublerisposta \ifx\srisposta\cartarisposta
\tothsize=\eudoublehsize \collhsize=\eucolumnsize
\vsize=\eudoublevsize  \else  \tothsize=\usdoublehsize
\collhsize=\uscolumnsize \vsize=\usdoublevsize \fi \else
\ifx\srisposta\cartarisposta \tothsize=\eusinglehsize
\vsize=\eusinglevsize \else  \tothsize=\ussinglehsize
\vsize=\ussinglevsize \fi \collhsize=4.4truein \fi
\ifx\risposta\mplarisposta \tothsize=5.0truein
\vsize=7.8truein \collhsize=4.4truein \fi
%
%
%
%
\newcount\contaeuler \newcount\contacyrill \newcount\contaams \newcount\contasym
\font\ninerm=cmr9  \font\eightrm=cmr8  \font\sixrm=cmr6
\font\ninei=cmmi9  \font\eighti=cmmi8  \font\sixi=cmmi6
\font\ninesy=cmsy9  \font\eightsy=cmsy8  \font\sixsy=cmsy6
\font\ninebf=cmbx9  \font\eightbf=cmbx8  \font\sixbf=cmbx6
\font\ninett=cmtt9  \font\eighttt=cmtt8  \font\nineit=cmti9
\font\eightit=cmti8 \font\ninesl=cmsl9  \font\eightsl=cmsl8
\skewchar\ninei='177 \skewchar\eighti='177 \skewchar\sixi='177
\skewchar\ninesy='60 \skewchar\eightsy='60 \skewchar\sixsy='60
\hyphenchar\ninett=-1 \hyphenchar\eighttt=-1 \hyphenchar\tentt=-1
\def\bfmath{\cmmib}                 
\font\tencmmib=cmmib10  \newfam\cmmibfam  \skewchar\tencmmib='177
\font\tencmbsy=cmbsy10  \newfam\cmbsyfam  \skewchar\tencmbsy='60
\def\scaps{\cmcsc}                 
\font\tencmcsc=cmcsc10  \newfam\cmcscfam
\ifnum\ingrandimento=1095 
 
\font\bfone=cmbx10 at 10.95pt

\font\capsone=cmcsc10 at 10.95pt 

\else  
 
\font\bfone=cmbx10 at 12pt

\font\capsone=cmcsc10 at 12pt 
\fi
\def\chapterfont#1{\xdef\ttaarr{#1}}
\def\sectionfont#1{\xdef\ppaarr{#1}}
\def\ttaarr{\bf}		
\def\ppaarr{\sl}		

%
%
%
\newfam\eufmfam \newfam\msamfam \newfam\msbmfam \newfam\eufbfam
\def\Loadeulerfonts{\global\contaeuler=1 \ifx\arisposta\amsrisposta
\font\teneufm=eufm10              
\font\eighteufm=eufm8 \font\nineeufm=eufm9 \font\sixeufm=eufm6
\font\seveneufm=eufm7  \font\fiveeufm=eufm5
\font\teneufb=eufb10              
\font\eighteufb=eufb8 \font\nineeufb=eufb9 \font\sixeufb=eufb6
\font\seveneufb=eufb7  \font\fiveeufb=eufb5
\font\teneurm=eurm10              
\font\eighteurm=eurm8 \font\nineeurm=eurm9
\font\teneurb=eurb10              
\font\eighteurb=eurb8 \font\nineeurb=eurb9
\font\teneusm=eusm10              
\font\eighteusm=eusm8 \font\nineeusm=eusm9
\font\teneusb=eusb10              
\font\eighteusb=eusb8 \font\nineeusb=eusb9
\else \def\eufm{\tt} \def\eufb{\tt} \def\eurm{\tt} \def\eurb{\tt}
\def\eusm{\tt} \def\eusb{\tt}    \fi}
\def\loadamsmath{\global\contaams=1 \ifx\arisposta\amsrisposta
\font\tenmsam=msam10 \font\ninemsam=msam9 \font\eightmsam=msam8
\font\sevenmsam=msam7 \font\sixmsam=msam6 \font\fivemsam=msam5
\font\tenmsbm=msbm10 \font\ninemsbm=msbm9 \font\eightmsbm=msbm8
\font\sevenmsbm=msbm7 \font\sixmsbm=msbm6 \font\fivemsbm=msbm5
\else \def\msbm{\bf} \fi \def\Bbb{\msbm} \def\symbl{\msam} \tenpoint}
\def\loadcyrill{\global\contacyrill=1 \ifx\arisposta\amsrisposta
\font\tenwncyr=wncyr10 \font\ninewncyr=wncyr9 \font\eightwncyr=wncyr8
\font\tenwncyb=wncyr10 \font\ninewncyb=wncyr9 \font\eightwncyb=wncyr8
\font\tenwncyi=wncyr10 \font\ninewncyi=wncyr9 \font\eightwncyi=wncyr8
\else \def\cyrill{\sl} \def\cyrilb{\sl} \def\cyrili{\sl} \fi\tenpoint}
\catcode`\@=11
\def\undefine#1{\let#1\undefined}
\def\newsymbol#1#2#3#4#5{\let\next@\relax
 \ifnum#2=\@ne\let\next@\msafam@\else
 \ifnum#2=\tw@\let\next@\msbfam@\fi\fi
 \mathchardef#1="#3\next@#4#5}
\def\mathhexbox@#1#2#3{\relax
 \ifmmode\mathpalette{}{\m@th\mathchar"#1#2#3}%
 \else\leavevmode\hbox{$\m@th\mathchar"#1#2#3$}\fi}
\def\hexnumber@#1{\ifcase#1 0\or 1\or 2\or 3\or 4\or 5\or 6\or 7\or 8\or 
9\or A\or B\or C\or D\or E\or F\fi}
\edef\msafam@{\hexnumber@\msamfam}
\edef\msbfam@{\hexnumber@\msbmfam}
\mathchardef\dabar@"0\msafam@39
\catcode`\@=12    
\def\loadamssym{\ifx\arisposta\amsrisposta  \ifnum\contaams=1 
\global\contasym=1 
\catcode`\@=11
\def\dashrightarrow{\mathrel{\dabar@\dabar@\mathchar"0\msafam@4B}}
\def\dashleftarrow{\mathrel{\mathchar"0\msafam@4C\dabar@\dabar@}}
\let\dasharrow\dashrightarrow
\def\ulcorner{\delimiter"4\msafam@70\msafam@70 }
\def\urcorner{\delimiter"5\msafam@71\msafam@71 }
\def\llcorner{\delimiter"4\msafam@78\msafam@78 }
\def\lrcorner{\delimiter"5\msafam@79\msafam@79 }
\def\yen{{\mathhexbox@\msafam@55}}
\def\checkmark{{\mathhexbox@\msafam@58 }}
\def\circledR{{\mathhexbox@\msafam@72 }}
\def\maltese{{\mathhexbox@\msafam@7A }}
\catcode`\@=12 
\input amssym.tex     \else  
\message{Panda error - First you have to use loadamsmath !!!!} \fi
\else \message{Panda error - You need the AMSFonts for these symbols 
!!!!}\fi}
\ifx\arisposta\amsrisposta
\font\sevenex=cmex7               
\font\eightex=cmex8  \font\nineex=cmex9
\font\ninecmmib=cmmib9   \font\eightcmmib=cmmib8
\font\sevencmmib=cmmib7 \font\sixcmmib=cmmib6
\font\fivecmmib=cmmib5   \skewchar\ninecmmib='177
\skewchar\eightcmmib='177  \skewchar\sevencmmib='177
\skewchar\sixcmmib='177   \skewchar\fivecmmib='177
\font\ninecmbsy=cmbsy9    \font\eightcmbsy=cmbsy8
\font\sevencmbsy=cmbsy7  \font\sixcmbsy=cmbsy6
\font\fivecmbsy=cmbsy5   \skewchar\ninecmbsy='60
\skewchar\eightcmbsy='60  \skewchar\sevencmbsy='60
\skewchar\sixcmbsy='60    \skewchar\fivecmbsy='60
\font\ninecmcsc=cmcsc9    \font\eightcmcsc=cmcsc8     \else
\def\cmmib{\fam\cmmibfam\tencmmib}\textfont\cmmibfam=\tencmmib
\scriptfont\cmmibfam=\tencmmib \scriptscriptfont\cmmibfam=\tencmmib
\def\cmbsy{\fam\cmbsyfam\tencmbsy} \textfont\cmbsyfam=\tencmbsy
\scriptfont\cmbsyfam=\tencmbsy \scriptscriptfont\cmbsyfam=\tencmbsy
\scriptfont\cmcscfam=\tencmcsc \scriptscriptfont\cmcscfam=\tencmcsc
\def\cmcsc{\fam\cmcscfam\tencmcsc} \textfont\cmcscfam=\tencmcsc \fi
\catcode`@=11
\newskip\ttglue
\gdef\tenpoint{\def\rm{\fam0\tenrm}
  \textfont0=\tenrm \scriptfont0=\sevenrm \scriptscriptfont0=\fiverm
  \textfont1=\teni \scriptfont1=\seveni \scriptscriptfont1=\fivei
  \textfont2=\tensy \scriptfont2=\sevensy \scriptscriptfont2=\fivesy
  \textfont3=\tenex \scriptfont3=\tenex \scriptscriptfont3=\tenex
  \def\mcal{\fam2 \tensy}  \def\mmit{\fam1 \teni}
  \textfont\itfam=\tenit \def\it{\fam\itfam\tenit}
  \textfont\slfam=\tensl \def\sl{\fam\slfam\tensl}
  \textfont\ttfam=\tentt \scriptfont\ttfam=\eighttt
  \scriptscriptfont\ttfam=\eighttt  \def\tt{\fam\ttfam\tentt}
  \textfont\bffam=\tenbf \scriptfont\bffam=\sevenbf
  \scriptscriptfont\bffam=\fivebf \def\bf{\fam\bffam\tenbf}
     \ifx\arisposta\amsrisposta    \ifnum\contaeuler=1
  \textfont\eufmfam=\teneufm \scriptfont\eufmfam=\seveneufm
  \scriptscriptfont\eufmfam=\fiveeufm \def\eufm{\fam\eufmfam\teneufm}
  \textfont\eufbfam=\teneufb \scriptfont\eufbfam=\seveneufb
  \scriptscriptfont\eufbfam=\fiveeufb \def\eufb{\fam\eufbfam\teneufb}
  \def\eurm{\teneurm} \def\eurb{\teneurb} \def\eusm{\teneusm}
  \def\eusb{\teneusb}    \fi    \ifnum\contaams=1
  \textfont\msamfam=\tenmsam \scriptfont\msamfam=\sevenmsam
  \scriptscriptfont\msamfam=\fivemsam \def\msam{\fam\msamfam\tenmsam}
  \textfont\msbmfam=\tenmsbm \scriptfont\msbmfam=\sevenmsbm
  \scriptscriptfont\msbmfam=\fivemsbm \def\msbm{\fam\msbmfam\tenmsbm}
     \fi      \ifnum\contacyrill=1     \def\cyrill{\tenwncyr}
  \def\cyrilb{\tenwncyb}  \def\cyrili{\tenwncyi}         \fi
  \textfont3=\tenex \scriptfont3=\sevenex \scriptscriptfont3=\sevenex
  \def\cmmib{\fam\cmmibfam\tencmmib} \scriptfont\cmmibfam=\sevencmmib
  \textfont\cmmibfam=\tencmmib  \scriptscriptfont\cmmibfam=\fivecmmib
  \def\cmbsy{\fam\cmbsyfam\tencmbsy} \scriptfont\cmbsyfam=\sevencmbsy
  \textfont\cmbsyfam=\tencmbsy  \scriptscriptfont\cmbsyfam=\fivecmbsy
  \def\cmcsc{\fam\cmcscfam\tencmcsc} \scriptfont\cmcscfam=\eightcmcsc
  \textfont\cmcscfam=\tencmcsc \scriptscriptfont\cmcscfam=\eightcmcsc
     \fi            \tt \ttglue=.5em plus.25em minus.15em
  \normalbaselineskip=12pt
  \setbox\strutbox=\hbox{\vrule height8.5pt depth3.5pt width0pt}
  \let\sc=\eightrm \let\big=\tenbig   \normalbaselines
  \baselineskip=\infralinea  \rm}
\gdef\ninepoint{\def\rm{\fam0\ninerm}
  \textfont0=\ninerm \scriptfont0=\sixrm \scriptscriptfont0=\fiverm
  \textfont1=\ninei \scriptfont1=\sixi \scriptscriptfont1=\fivei
  \textfont2=\ninesy \scriptfont2=\sixsy \scriptscriptfont2=\fivesy
  \textfont3=\tenex \scriptfont3=\tenex \scriptscriptfont3=\tenex
  \def\mcal{\fam2 \ninesy}  \def\mmit{\fam1 \ninei}
  \textfont\itfam=\nineit \def\it{\fam\itfam\nineit}
  \textfont\slfam=\ninesl \def\sl{\fam\slfam\ninesl}
  \textfont\ttfam=\ninett \scriptfont\ttfam=\eighttt
  \scriptscriptfont\ttfam=\eighttt \def\tt{\fam\ttfam\ninett}
  \textfont\bffam=\ninebf \scriptfont\bffam=\sixbf
  \scriptscriptfont\bffam=\fivebf \def\bf{\fam\bffam\ninebf}
     \ifx\arisposta\amsrisposta  \ifnum\contaeuler=1
  \textfont\eufmfam=\nineeufm \scriptfont\eufmfam=\sixeufm
  \scriptscriptfont\eufmfam=\fiveeufm \def\eufm{\fam\eufmfam\nineeufm}
  \textfont\eufbfam=\nineeufb \scriptfont\eufbfam=\sixeufb
  \scriptscriptfont\eufbfam=\fiveeufb \def\eufb{\fam\eufbfam\nineeufb}
  \def\eurm{\nineeurm} \def\eurb{\nineeurb} \def\eusm{\nineeusm}
  \def\eusb{\nineeusb}     \fi   \ifnum\contaams=1
  \textfont\msamfam=\ninemsam \scriptfont\msamfam=\sixmsam
  \scriptscriptfont\msamfam=\fivemsam \def\msam{\fam\msamfam\ninemsam}
  \textfont\msbmfam=\ninemsbm \scriptfont\msbmfam=\sixmsbm
  \scriptscriptfont\msbmfam=\fivemsbm \def\msbm{\fam\msbmfam\ninemsbm}
     \fi       \ifnum\contacyrill=1     \def\cyrill{\ninewncyr}
  \def\cyrilb{\ninewncyb}  \def\cyrili{\ninewncyi}         \fi
  \textfont3=\nineex \scriptfont3=\sevenex \scriptscriptfont3=\sevenex
  \def\cmmib{\fam\cmmibfam\ninecmmib}  \textfont\cmmibfam=\ninecmmib
  \scriptfont\cmmibfam=\sixcmmib \scriptscriptfont\cmmibfam=\fivecmmib
  \def\cmbsy{\fam\cmbsyfam\ninecmbsy}  \textfont\cmbsyfam=\ninecmbsy
  \scriptfont\cmbsyfam=\sixcmbsy \scriptscriptfont\cmbsyfam=\fivecmbsy
  \def\cmcsc{\fam\cmcscfam\ninecmcsc} \scriptfont\cmcscfam=\eightcmcsc
  \textfont\cmcscfam=\ninecmcsc \scriptscriptfont\cmcscfam=\eightcmcsc
     \fi            \tt \ttglue=.5em plus.25em minus.15em
  \normalbaselineskip=11pt
  \setbox\strutbox=\hbox{\vrule height8pt depth3pt width0pt}
  \let\sc=\sevenrm \let\big=\ninebig \normalbaselines\rm}
\gdef\eightpoint{\def\rm{\fam0\eightrm}
  \textfont0=\eightrm \scriptfont0=\sixrm \scriptscriptfont0=\fiverm
  \textfont1=\eighti \scriptfont1=\sixi \scriptscriptfont1=\fivei
  \textfont2=\eightsy \scriptfont2=\sixsy \scriptscriptfont2=\fivesy
  \textfont3=\tenex \scriptfont3=\tenex \scriptscriptfont3=\tenex
  \def\mcal{\fam2 \eightsy}  \def\mmit{\fam1 \eighti}
  \textfont\itfam=\eightit \def\it{\fam\itfam\eightit}
  \textfont\slfam=\eightsl \def\sl{\fam\slfam\eightsl}
  \textfont\ttfam=\eighttt \scriptfont\ttfam=\eighttt
  \scriptscriptfont\ttfam=\eighttt \def\tt{\fam\ttfam\eighttt}
  \textfont\bffam=\eightbf \scriptfont\bffam=\sixbf
  \scriptscriptfont\bffam=\fivebf \def\bf{\fam\bffam\eightbf}
     \ifx\arisposta\amsrisposta   \ifnum\contaeuler=1
  \textfont\eufmfam=\eighteufm \scriptfont\eufmfam=\sixeufm
  \scriptscriptfont\eufmfam=\fiveeufm \def\eufm{\fam\eufmfam\eighteufm}
  \textfont\eufbfam=\eighteufb \scriptfont\eufbfam=\sixeufb
  \scriptscriptfont\eufbfam=\fiveeufb \def\eufb{\fam\eufbfam\eighteufb}
  \def\eurm{\eighteurm} \def\eurb{\eighteurb} \def\eusm{\eighteusm}
  \def\eusb{\eighteusb}       \fi    \ifnum\contaams=1
  \textfont\msamfam=\eightmsam \scriptfont\msamfam=\sixmsam
  \scriptscriptfont\msamfam=\fivemsam \def\msam{\fam\msamfam\eightmsam}
  \textfont\msbmfam=\eightmsbm \scriptfont\msbmfam=\sixmsbm
  \scriptscriptfont\msbmfam=\fivemsbm \def\msbm{\fam\msbmfam\eightmsbm}
     \fi       \ifnum\contacyrill=1     \def\cyrill{\eightwncyr}
  \def\cyrilb{\eightwncyb}  \def\cyrili{\eightwncyi}         \fi
  \textfont3=\eightex \scriptfont3=\sevenex \scriptscriptfont3=\sevenex
  \def\cmmib{\fam\cmmibfam\eightcmmib}  \textfont\cmmibfam=\eightcmmib
  \scriptfont\cmmibfam=\sixcmmib \scriptscriptfont\cmmibfam=\fivecmmib
  \def\cmbsy{\fam\cmbsyfam\eightcmbsy}  \textfont\cmbsyfam=\eightcmbsy
  \scriptfont\cmbsyfam=\sixcmbsy \scriptscriptfont\cmbsyfam=\fivecmbsy
  \def\cmcsc{\fam\cmcscfam\eightcmcsc} \scriptfont\cmcscfam=\eightcmcsc
  \textfont\cmcscfam=\eightcmcsc \scriptscriptfont\cmcscfam=\eightcmcsc
     \fi             \tt \ttglue=.5em plus.25em minus.15em
  \normalbaselineskip=9pt
  \setbox\strutbox=\hbox{\vrule height7pt depth2pt width0pt}
  \let\sc=\sixrm \let\big=\eightbig \normalbaselines\rm }
\gdef\tenbig#1{{\hbox{$\left#1\vbox to8.5pt{}\right.\n@space$}}}
\gdef\ninebig#1{{\hbox{$\textfont0=\tenrm\textfont2=\tensy
   \left#1\vbox to7.25pt{}\right.\n@space$}}}
\gdef\eightbig#1{{\hbox{$\textfont0=\ninerm\textfont2=\ninesy
   \left#1\vbox to6.5pt{}\right.\n@space$}}}
\def\alternativefont#1#2{\ifx\arisposta\amsrisposta \relax \else
\xdef#1{#2} \fi}
\global\contaeuler=0 \global\contacyrill=0 \global\contaams=0
%
%
%
%
\newbox\fotlinebb \newbox\hedlinebb \newbox\leftcolumn
\gdef\makeheadline{\vbox to 0pt{\vskip-22.5pt
     \fullline{\vbox to8.5pt{}\the\headline}\vss}\nointerlineskip}
\gdef\makehedlinebb{\vbox to 0pt{\vskip-22.5pt
     \fullline{\vbox to8.5pt{}\copy\hedlinebb\hfil
     \line{\hfill\the\headline\hfill}}\vss} \nointerlineskip}
\gdef\makefootline{\baselineskip=24pt \fullline{\the\footline}}
\gdef\makefotlinebb{\baselineskip=24pt
    \fullline{\copy\fotlinebb\hfil\line{\hfill\the\footline\hfill}}}
\gdef\doubleformat{\shipout\vbox{\Landspec\makehedlinebb
     \fullline{\box\leftcolumn\hfil\columnbox}\makefotlinebb}
     \advancepageno}
\gdef\columnbox{\leftline{\pagebody}}
\gdef\line#1{\hbox to\hsize{\hskip\leftskip#1\hskip\rightskip}}
\gdef\fullline#1{\hbox to\fullhsize{\hskip\leftskip{#1}%
\hskip\rightskip}}
\gdef\footnote#1{\let\@sf=\empty
         \ifhmode\edef\#sf{\spacefactor=\the\spacefactor}\/\fi
         #1\@sf\vfootnote{#1}}
\gdef\vfootnote#1{\insert\footins\bgroup
         \ifnum\dimnota=1  \eightpoint\fi
         \ifnum\dimnota=2  \ninepoint\fi
         \ifnum\dimnota=0  \tenpoint\fi
         \interlinepenalty=\interfootnotelinepenalty
         \splittopskip=\ht\strutbox
         \splitmaxdepth=\dp\strutbox \floatingpenalty=20000
         \leftskip=\oldssposta \rightskip=\olddsposta
         \spaceskip=0pt \xspaceskip=0pt
         \ifnum\sinnota=0   \textindent{#1}\fi
         \ifnum\sinnota=1   \item{#1}\fi
         \footstrut\futurelet\next\fo@t}
\gdef\fo@t{\ifcat\bgroup\noexpand\next \let\next\f@@t
             \else\let\next\f@t\fi \next}
\gdef\f@@t{\bgroup\aftergroup\@foot\let\next}
\gdef\f@t#1{#1\@foot} \gdef\@foot{\strut\egroup}
\gdef\footstrut{\vbox to\splittopskip{}}
\skip\footins=\bigskipamount
\count\footins=1000  \dimen\footins=8in
\catcode`@=12
\tenpoint
\ifnum\unoduecol=1 \hsize=\tothsize   \fullhsize=\tothsize \fi
\ifnum\unoduecol=2 \hsize=\collhsize  \fullhsize=\tothsize \fi
\global\let\lrcol=L      \ifnum\unoduecol=1
\output{\plainoutput{\ifnum\tipbnota=2 \clearnmbnota\fi}} \fi
\ifnum\unoduecol=2 \output{\if L\lrcol
     \global\setbox\leftcolumn=\columnbox
     \global\setbox\fotlinebb=\line{\hfill\the\footline\hfill}
     \global\setbox\hedlinebb=\line{\hfill\the\headline\hfill}
     \advancepageno  \global\let\lrcol=R
     \else  \doubleformat \global\let\lrcol=L \fi
     \ifnum\outputpenalty>-20000 \else\dosupereject\fi
     \ifnum\tipbnota=2\clearnmbnota\fi }\fi
\def\ifdoublepage{\ifnum\unoduecol=2 }
\gdef\yespagenumbers{\footline={\hss\tenrm\folio\hss}}
\gdef\ciao{ \ifnum\fdefcontre=1 \endfdef\fi
     \par\vfill\supereject \ifnum\unoduecol=2
     \if R\lrcol  \headline={}\nopagenumbers\null\vfill\eject
     \fi\fi \end}

\newskip\olddsposta \newskip\oldssposta
\global\oldssposta=\leftskip \global\olddsposta=\rightskip

\def\filldots{\leaders\hbox to 1em{\hss.\hss}\hfill}
\def\inquadrb#1 {\vbox {\hrule  \hbox{\vrule \vbox {\vskip .2cm
    \hbox {\ #1\ } \vskip .2cm } \vrule  }  \hrule} }
 \def\newline{\hfil\break}
\def\jump{\vskip\baselineskip} \newskip\iinnffrr
\def\sjump{\iinnffrr=\baselineskip
          \divide\iinnffrr by 2 \vskip\iinnffrr}
\def\bjump{\vskip\baselineskip \vskip\baselineskip}
\newcount\nmbnota  \def\clearnmbnota{\global\nmbnota=0}
\newcount\tipbnota \def\letterfootnote{\global\tipbnota=1}

\def\note#1{\global\advance\nmbnota by 1 \ifnum\tipbnota=1
    \footnote{$^{\rm\nttlett}$}{#1} \else {\ifnum\tipbnota=2
    \footnote{$^{\nttsymb}$}{#1}
    \else\footnote{$^{\the\nmbnota}$}{#1}\fi}\fi}
\def\nttlett{\ifcase\nmbnota \or a\or b\or c\or d\or e\or f\or
g\or h\or i\or j\or k\or l\or m\or n\or o\or p\or q\or r\or
s\or t\or u\or v\or w\or y\or x\or z\fi}
\def\nttsymb{\ifcase\nmbnota \or\dag\or\sharp\or\ddag\or\star\or
\natural\or\flat\or\clubsuit\or\diamondsuit\or\heartsuit
\or\spadesuit\fi}   \clearnmbnota
\def\numberfootnote{\global\tipbnota=0} \numberfootnote
\def\setnote#1{\expandafter\xdef\csname#1\endcsname{
\ifnum\tipbnota=1 {\rm\nttlett} \else {\ifnum\tipbnota=2
{\nttsymb} \else \the\nmbnota\fi}\fi} }
\newcount\nbmfig  \def\clearnbmfig{\global\nbmfig=0}
\gdef\figure{\global\advance\nbmfig by 1
      {\rm fig. \the\nbmfig}}   \clearnbmfig
\def\setfig#1{\expandafter\xdef\csname#1\endcsname{fig. \the\nbmfig}}
 \def\endformula{\eqno\numero $$}
 \def\efr{\endformula}
\newcount\frmcount \def\clearfrmcount{\global\frmcount=0}
\def\numero{\global\advance\frmcount by 1   \ifnum\indappcount=0
  {\ifnum\cpcount <1 {\hbox{\rm (\the\frmcount )}}  \else
  {\hbox{\rm (\the\cpcount .\the\frmcount )}} \fi}  \else
  {\hbox{\rm (\applett .\the\frmcount )}} \fi}
\def\nfr{\nameformula}    \def\numali{\numero}
\def\nameformula#1{\global\advance\frmcount by 1%
{\ifnum\indappcount=0%
{\ifnum\cpcount<1\xdef\spzzttrra{(\the\frmcount )}%
\else\xdef\spzzttrra{(\the\cpcount .\the\frmcount )}\fi}%
\else\xdef\spzzttrra{(\applett .\the\frmcount )}\fi}%
\expandafter\xdef\csname#1\endcsname{\spzzttrra}%
\eqno{\ifnum\draftnum=0\hbox{\rm\spzzttrra}\else%
\hbox{$\buildchar{\rm\spzzttrra}{\tt\scriptscriptstyle#1}{}$}\fi}$$}
\def\nameali#1{\global\advance\frmcount by 1%
{\ifnum\indappcount=0%
{\ifnum\cpcount<1\xdef\spzzttrra{(\the\frmcount )}%
\else\xdef\spzzttrra{(\the\cpcount .\the\frmcount )}\fi}%
\else\xdef\spzzttrra{(\applett .\the\frmcount )}\fi}%
\expandafter\xdef\csname#1\endcsname{\spzzttrra}%
\ifnum\draftnum=0\hbox{\rm\spzzttrra}\else%
\hbox{$\buildchar{\rm\spzzttrra}{\tt\scriptscriptstyle#1}{}$}\fi}
\clearfrmcount
\newcount\cpcount \def\clearcpcount{\global\cpcount=0}
\newcount\subcpcount \def\clearsubcpcount{\global\subcpcount=0}
\newcount\appcount \def\clearappcount{\global\appcount=0}
\newcount\indappcount \def\clearindappcount{\indappcount=0}
\newcount\sottoparcount 

\def\applett{\ifcase\appcount  \or {A}\or {B}\or {C}\or
{D}\or {E}\or {F}\or {G}\or {H}\or {I}\or {J}\or {K}\or {L}\or
{M}\or {N}\or {O}\or {P}\or {Q}\or {R}\or {S}\or {T}\or {U}\or
{V}\or {W}\or {X}\or {Y}\or {Z}\fi    \ifnum\appcount<0
\immediate\write16 {Panda ERROR - Appendix: counter "appcount"
out of range}\fi  \ifnum\appcount>26  \immediate\write16 {Panda
ERROR - Appendix: counter "appcount" out of range}\fi}
\clearappcount  \clearindappcount \newcount\connttrre
\def\clearconnttrre{\global\connttrre=0} \newcount\countref
\def\clearcountref{\global\countref=0} \clearcountref
\def\chapter#1{\global\advance\cpcount by 1 \clearfrmcount
                 \goodbreak\null\vbox{\jump\nobreak
                 \clearsubcpcount\clearindappcount
                 \itemitem{\ttaarr\the\cpcount .\qquad}{\ttaarr #1}
                 \par\nobreak\jump\sjump}\nobreak}
\def\section#1{\global\advance\subcpcount by 1 \goodbreak\null
               \vbox{\sjump\nobreak\ifnum\indappcount=0
                 {\ifnum\cpcount=0 {\itemitem{\ppaarr
               .\the\subcpcount\quad\enskip\ }{\ppaarr #1}\par} \else
                 {\itemitem{\ppaarr\the\cpcount .\the\subcpcount\quad
                  \enskip\ }{\ppaarr #1} \par}  \fi}
                \else{\itemitem{\ppaarr\applett .\the\subcpcount\quad
                 \enskip\ }{\ppaarr #1}\par}\fi\nobreak\jump}\nobreak}
\clearsubcpcount
\def\appendix#1{\global\advance\appcount by 1 \clearfrmcount
                  \goodbreak\null\vbox{\jump\nobreak
                  \global\advance\indappcount by 1 \clearsubcpcount
          \itemitem{ }{\hskip-40pt\ttaarr Appendix\ \applett :\ #1}
             \nobreak\jump\sjump}\nobreak}
\clearappcount \clearindappcount
\def\references{\goodbreak\null\vbox{\jump\nobreak
   \itemitem{}{\ttaarr References} \nobreak\jump\sjump}\nobreak}

\def\introsumm{\clearindappcount\clearappcount\clearcpcount
                  \clearsubcpcount\goodbreak\null\vbox{\jump\nobreak
  \itemitem{}{\ttaarr Introduction and Summary} \nobreak\jump\sjump}\nobreak}
\clearcpcount\clearcountref

\def\setchap#1{\ifnum\indappcount=0{\ifnum\subcpcount=0%
\xdef\spzzttrra{\the\cpcount}%
\else\xdef\spzzttrra{\the\cpcount .\the\subcpcount}\fi}
\else{\ifnum\subcpcount=0 \xdef\spzzttrra{\applett}%
\else\xdef\spzzttrra{\applett .\the\subcpcount}\fi}\fi
\expandafter\xdef\csname#1\endcsname{\spzzttrra}}
\newcount\draftnum \newcount\ppora   \newcount\ppminuti
\global\ppora=\time   \global\ppminuti=\time
\global\divide\ppora by 60  \draftnum=\ppora
\multiply\draftnum by 60    \global\advance\ppminuti by -\draftnum
\def\droggi{\number\day /\number\month /\number\year\ \the\ppora
:\the\ppminuti}     \global\draftnum=0
\def\draftcomment#1{\ifnum\draftnum=0 \relax \else
{\ {\bf ***}\ #1\ {\bf ***}\ }\fi} 
%
%
\catcode`@=11
\gdef\Ref#1{\expandafter\ifx\csname @rrxx@#1\endcsname\relax%
{\global\advance\countref by 1    \ifnum\countref>200
\immediate\write16 {Panda ERROR - Ref: maximum number of references
exceeded}  \expandafter\xdef\csname @rrxx@#1\endcsname{0}\else
\expandafter\xdef\csname @rrxx@#1\endcsname{\the\countref}\fi}\fi
\ifnum\draftnum=0 \csname @rrxx@#1\endcsname \else#1\fi}
\gdef\beginref{\ifnum\draftnum=0  \gdef\Rref{\fairef}
\gdef\endref{\scriviref} \else\relax\fi
\ifx\risposta\mplarisposta \ninepoint \fi
\baselineskip=12pt \parskip 2pt plus.2pt }
\def\Reflab#1{[#1]} \gdef\Rref#1#2{\item{\Reflab{#1}}{#2}}
\gdef\endref{\relax}  \newcount\conttemp
\gdef\fairef#1#2{\expandafter\ifx\csname @rrxx@#1\endcsname\relax
{\global\conttemp=0 \immediate\write16 {Panda ERROR - Ref: reference
[#1] undefined}} \else
{\global\conttemp=\csname @rrxx@#1\endcsname } \fi
\global\advance\conttemp by 50  \global\setbox\conttemp=\hbox{#2} }
\gdef\scriviref{\clearconnttrre\conttemp=50
\loop\ifnum\connttrre<\countref \advance\conttemp by 1
\advance\connttrre by 1
\item{\Reflab{\the\connttrre}}{\unhcopy\conttemp} \repeat}
\clearcountref \clearconnttrre
\catcode`@=12
\ifx\risposta\mplarisposta \def\Reflab#1{#1.} \letterfootnote \fi
%
%

\def\slashchar#1{\setbox0=\hbox{$#1$} \dimen0=\wd0
     \setbox1=\hbox{/} \dimen1=\wd1 \ifdim\dimen0>\dimen1
      \rlap{\hbox to \dimen0{\hfil/\hfil}} #1 \else
      \rlap{\hbox to \dimen1{\hfil$#1$\hfil}} / \fi}
\ifx\oldchi\undefined \let\oldchi=\chi
  \def\cchi{{\raise 1pt\hbox{$\oldchi$}}} \let\chi=\cchi \fi
\ifnum\contasym=1 \else \fi

\def\frac#1#2{{\textstyle{#1 \over #2}}}

\def\half{\ifinner {\scriptstyle {1 \over 2}}\else {1 \over 2} \fi}
  \def\ket#1{\vert#1\rangle}

\def\vev#1{\langle#1\rangle}

\def\simge{\rlap{\raise 2pt \hbox{$>$}}{\lower 2pt \hbox{$\sim$}}}
\def\simle{\rlap{\raise 2pt \hbox{$<$}}{\lower 2pt \hbox{$\sim$}}}

\def\buildchar#1#2#3{{\null\!\mathop{#1}\limits^{#2}_{#3}\!\null}}

\def\vbig#1#2{{\vbigd@men=#2\divide\vbigd@men by 2%
\hbox{$\left#1\vbox to \vbigd@men{}\right.\n@space$}}}

\def\noblackbox{\overfullrule=0pt} 
%
%
\newcount\fdefcontre \newcount\fdefcount \newcount\indcount
\newread\filefdef  \newread\fileftmp  \newwrite\filefdef
\newwrite\fileftmp     \def\strip #1*.A {#1}%
\def\futuredef#1{\beginfdef
\expandafter\ifx\csname#1\endcsname\relax%
{\immediate\write\fileftmp{#1*.A}%
\immediate\write16 {Panda Warning - fdef: macro "#1" on page
\the\pageno \space undefined}
\ifnum\draftnum=0 \expandafter\xdef\csname#1\endcsname{(?)}
\else \expandafter\xdef\csname#1\endcsname{(#1)}\fi
\global\advance\fdefcount by 1}\fi\csname#1\endcsname}

\def\beginfdef{\ifnum\fdefcontre=0
\immediate\openin\filefdef\jobname.fdef
\immediate\openout\fileftmp\jobname.ftmp
\global\fdefcontre=1  \ifeof\filefdef \immediate\write16 {Panda
WARNING - fdef: file \jobname.fdef not found, run TeX again}
\else \immediate\read\filefdef to\spzzttrra
\global\advance\fdefcount by \spzzttrra
\indcount=0 \loop\ifnum\indcount<\fdefcount
\advance\indcount by 1%
\immediate\read\filefdef to\spezttrra%
\immediate\read\filefdef to\sppzttrra%
\edef\spzzttrra{\expandafter\strip\spezttrra}%
\immediate\write\fileftmp {\spzzttrra *.A}
\expandafter\xdef\csname\spzzttrra\endcsname{\sppzttrra}%
\repeat \fi \immediate\closein\filefdef \fi}
\def\endfdef{\immediate\closeout\fileftmp   \ifnum\fdefcount>0
\immediate\openin\fileftmp \jobname.ftmp
\immediate\openout\filefdef \jobname.fdef
\immediate\write\filefdef {\the\fdefcount}   \indcount=0
\loop\ifnum\indcount<\fdefcount    \advance\indcount by 1
\immediate\read\fileftmp to\spezttrra
\edef\spzzttrra{\expandafter\strip\spezttrra}
\immediate\write\filefdef{\spzzttrra *.A}
\edef\spezttrra{\string{\csname\spzzttrra\endcsname\string}}
\iwritel\filefdef{\spezttrra}
\repeat  \immediate\closein\fileftmp \immediate\closeout\filefdef
\immediate\write16 {Panda Warning - fdef: Label(s) may have changed,
re-run TeX to get them right}\fi}
\def\iwritel#1#2{\newlinechar=-1
{\newlinechar=`\ \immediate\write#1{#2}}\newlinechar=-1}
\global\fdefcontre=0 \global\fdefcount=0 \global\indcount=0
%
%
%
\mathchardef\alpha="710B   \mathchardef\beta="710C
\mathchardef\gamma="710D   \mathchardef\delta="710E
\mathchardef\epsilon="710F   \mathchardef\zeta="7110
\mathchardef\eta="7111   \mathchardef\theta="7112
\mathchardef\iota="7113   \mathchardef\kappa="7114
\mathchardef\lambda="7115   \mathchardef\mu="7116
\mathchardef\nu="7117   \mathchardef\xi="7118
\mathchardef\pi="7119   \mathchardef\rho="711A
\mathchardef\sigma="711B   \mathchardef\tau="711C
\mathchardef\upsilon="711D   \mathchardef\phi="711E
\mathchardef\chi="711F   \mathchardef\psi="7120
\mathchardef\omega="7121   \mathchardef\varepsilon="7122
\mathchardef\vartheta="7123   \mathchardef\varpi="7124
\mathchardef\varrho="7125   \mathchardef\varsigma="7126
\mathchardef\varphi="7127
%
%
\null
%
%
%
%

%
%
%
\loadamsmath 
\chapterfont{\bfone} \sectionfont{\scaps}
\noblackbox

\def\Teta#1#2{\Theta\left[{}^{#1}_{#2}\right]}

\def\wcptarrow{\ \buildchar{\longrightarrow}{{\rm{\scriptscriptstyle 
              WS-CPT}}}{ }\ }
\def\srarrow{\ \buildchar{\longrightarrow}{{\rm{\scriptscriptstyle 
              SR}}}{ }\ }
\def\bpzarrow{\ \buildchar{\longrightarrow}{{\rm{\scriptscriptstyle 
              BPZ}}}{ }\ }
\def\congphi{\ \buildchar{\cong}{{{\scriptscriptstyle 
              \Phi}}}{ }\ }

\def\eqhc{\ \buildchar{=}{\rm\scriptscriptstyle HC}{ }\ } 
\def\eqbpz{\ \buildchar{=}{\rm\scriptscriptstyle BPZ}{ }\ } 
\def\eqcpt{\ \buildchar{=}{\rm\scriptscriptstyle CPT}{ }\ } 
\def\eqwcpt{\ \buildchar{=}{\rm\scriptscriptstyle WS-CPT}{ }\ }

\def\di{{\rm d}}
\nopagenumbers
{\baselineskip=12pt
\line{\hfill IFUM-534/FT}
\line{\hfill NBI-HE-96-32}
\line{\hfill hep-th/9608022}
\line{\hfill August, 1996}}
{\baselineskip=14pt
\vfill
\centerline{\capsone String Theory and the CPT Theorem}
\sjump
\centerline{\capsone on the World-Sheet}
\bjump\bjump
\centerline{\scaps Andrea Pasquinucci~\footnote{$^\dagger$}{Supported
by MURST and by EU Science Programs  no.\ SC1*-CT92-0789 and
no.\ CHRX-CT-920035.} }
\sjump
\centerline{\sl Dipartimento di Fisica, Universit\`a di Milano}
\centerline{\sl and INFN, sezione di Milano}
\centerline{\sl via Celoria 16, I-20133 Milano, Italy}
\jump
\centerline{\rm and}
\jump
\centerline{\scaps Kaj Roland~\footnote{$^\ddagger$}{Supported by the 
Carlsberg Foundation and by EU Science Program  no.\ SC1*-CT92-0789.} }
\sjump
\centerline{\sl The Niels Bohr Institute, University of Copenhagen,}
\centerline{\sl Blegdamsvej 17, DK-2100, Copenhagen, Denmark}
\bjump \vfill
\ifdoublepage \eject\null\vfill\fi
\centerline{\capsone ABSTRACT}
\sjump
\noindent
We study the CPT theorem for a two-dimensional conformal field theory on 
an arbitrary Riemann surface. On the sphere the theorem follows from the 
assumption that the correlation functions have standard hermiticity 
properties and are invariant under the transformation $z \rightarrow 
1/z$. The theorem can then be extended to higher genus surfaces by 
sewing. We show that, as a consequence of the CPT theorem on the 
world-sheet, the scattering $T$-matrix in string theory is {\sl formally\/} 
hermitean at any loop order.

\sjump \vfill 
%
\pageno=0 \eject }
\yespagenumbers\pageno=1
%
\null\bjump
\introsumm
The CPT theorem is a very fundamental result which holds in any quantum 
field theory under very mild assumptions, notably Lorentz invariance and 
locality [\Ref{Pauli},\Ref{Luders},\Ref{others},\Ref{SW}]. 
In this paper we consider the role played by the CPT theorem 
in the framework of two-dimensional conformal field theory (CFT) [\Ref{BPZ}]. 

Our main interest is in those CFTs that enter into the construction 
of string theory models; 
accordingly, we would like to study two-dimensional CFT on a compact 
Riemann surface of any genus, rather than in flat two-dimensional 
Minkowski space-time. This means that we loose Lorentz invariance, one 
of the main pillars of the ordinary CPT theorem, from the very outset. 
Nevertheless, as we shall see, this is more than compensated for by the 
presence of the much more powerful conformal symmetry. Indeed, the twin 
assumptions of conformal invariance (more precisely, invariance under 
the transformation $z \rightarrow 1/z$) and hermiticity (to be 
formulated more precisely in section 2) leads to the identity between 
correlation functions on the sphere which is usually referred to as 
CPT invariance of two-dimensional CFT~[\Ref{MooreSeiberg},\Ref{Sonoda}].

By using the sewing procedure of Sonoda [\Ref{Sonodatwo}] one may then try to
extend the concept of world-sheet CPT (WS-CPT) invariance from the sphere
to higher genus Riemann surfaces, where the global transformation 
$z \rightarrow 1/z$ ceases to be well-defined. 

We start by considering local CFTs, i.e. theories where the difference 
between left- and right-moving conformal dimensions is 
always integer. If one further assumes modular invariance of the correlation 
functions defined by the sewing (plus a very mild technical
requirement specified in section 3) the WS-CPT theorem can be extended
to higher genus Riemann surfaces. At genus $g \geq 1$, WS-CPT
invariance relates correlation functions on {\it different} Riemann 
surfaces, $M$ and $\tilde{M}$, that 
can be thought of as ``mirror images'' of one another. 

Although the {\it full} CFT describing a consistent string theory is 
always local, this is not necessarily true of the various building blocks 
involved in the construction of the string model; important examples are 
free complex fermions, free chiral bosons, and the superghosts. 
We therefore proceed to consider the 
modifications needed in the formulation of WS-CPT invariance in these 
specific examples. 
In particular, we show how ordinary WS-CPT invariance, in the 
sense of local, modular invariant CFT, is recovered when we put the 
various ``constituent'' CFTs together in a consistent string model and
sum over the spin structures.

It is also possible to define a WS-CPT transformation acting on 
individual operators. Unlike the ordinary CPT transformation in 
Minkowski space-time, the WS-CPT transformation inverts the order of 
operators and therefore cannot itself be generated by any operator. 
In the context of string theory, the WS-CPT transformation maps 
the vertex operator of an incoming physical string state into the vertex 
operator describing the {\it same} physical string state, but outgoing, 
times a certain phase factor [\Ref{normaliz},\Ref{uscpt}].

The CFTs we consider may, of course, be super-conformal, since any
super-conformal field theory is also conformal. In the case of
super-conformal field theories the natural aim would be to define WS-CPT on
arbitrary {\it super} Riemann surfaces. Instead, we consider only ordinary
Riemann surfaces, the reason being that we are mainly interested in string
theory, where we always imagine the supermoduli to be integrated out.
As is well known~[\Ref{Verlinde}], 
this gives rise to scattering amplitudes expressed
as integrals over the moduli space of ordinary Riemann surfaces, where
the integrand is given by a correlation function of vertex operators
with an appropriate number of Picture Changing Operators (PCOs)
inserted.

When applied to the CFT underlying a given string theory, WS-CPT 
invariance implies that the scattering $T$-matrix is hermitean at 
tree-level away from the momentum poles, as required by unitarity.
More generally, we show that the $T$-matrix is formally hermitean at any 
loop order, meaning that it is hermitean to the extent that the integral 
over the moduli is convergent. Thus, if we believe in unitarity of the 
scattering $S$-matrix in string theory, our result can be interpreted as 
an indirect demonstration that the modular integral can not be 
convergent in those kinematical regions where the $T$-matrix is required 
to develop an imaginary part. Indeed, unitarity becomes a consistency 
condition that should be imposed on whatever regularization is 
introduced to handle the divergences. Several regularization
procedures (of varying generality) have already appeared in the 
literature~[\Ref{Amano},\Ref{Mitchell},\Ref{Hoker},\Ref{Weisberger},\Ref{Berera}].

The concept of WS-CPT invariance in two-dimensional CFT was first introduced 
by Moore and Seiberg as one of the fundamental assumptions underlying
their axiomatic discussion of CFT [\Ref{MooreSeiberg}]. 

Sonoda [\Ref{Sonoda}] extended the property of WS-CPT invariance for
local and unitary CFTs to higher genus Riemann surfaces by means of 
sewing, and he also noticed the connection between WS-CPT invariance 
and the hermiticity of the (dispersive part of the) $T$-matrix in the 
context of string theory. 

Compared to these authors, our point of view is somewhat different:
WS-CPT invariance on the sphere is not considered to be an axiom; instead,
it is seen as a consequence of the twin assumptions of 
conformal invariance and hermiticity. 
We do not restrict ourselves to {\it unitary\/} CFTs; 
all we assume is that we have an inner product defined on the Hilbert space, 
and hence a concept of hermitean conjugation of all operators, such that 
the energy-momentum tensor of the CFT is hermitean. The inner product 
is {\it not} assumed to be positive definite --- states may have 
zero or negative norm. 
Therefore we include such important non-unitary theories as the 
time-like component of the space-time coordinate in 
string theory, and the reparametrization ghosts. 

Another new feature of our paper is the discussion of world-sheet CPT
for free non-local CFTs such as complex fermions and superghosts.

The paper is organized as follows:
In section 1 we briefly review the CPT theorem for quantum 
field theories in $D$-dimensional Minkowski space-time. Then,
in section 2, we discuss the CPT theorem for CFTs on the 
sphere, including the question of hermiticity.
In section 3 we extend the CPT theorem to higher 
genus Riemann surfaces by means of the sewing technique, first for local 
CFTs, then for some important examples of free non-local CFTs.
Finally, in section 4 we apply the previous results to the CFT 
underlying any given first-quantized string theory and we show that 
the CPT theorem on the world-sheet implies the {\sl formal\/} hermiticity of 
the $T$-matrix amplitudes to all orders in string perturbation theory.
We also provide an Appendix on mirror image Riemann surfaces.
\chapter{CPT theorem in $D$ dimensional Minkowski space-time.}
In this section we briefly review the CPT theorem and 
CPT transformations in even-dimensional 
Min\-kow\-ski space-time in the context of quantum field theory, mostly to fix 
our notations (see ref. [\Ref{uscpt}] for more details).

The CPT theorem [\Ref{Pauli},\Ref{Luders},\Ref{others},\Ref{SW}] 
asserts that any quantum field theory is 
invariant under CPT transformations assuming that it satisfies the 
following very mild assumptions:
a) Lorentz invariance,
b) The energy is positive definite and there exists a 
Poincar\'{e}-invariant vacuum, unique up to a phase factor,
c) Local commutativity, i.e. field operators at space-like 
separations either commute or anti-commute.
These assumptions also imply the spin-statistics theorem, i.e. fields of 
integer (half odd integer) spin~\note{In $D$ dimensions a field is
said to have integer (half odd integer) spin if it furnishes a
representation of the Lorentz algebra $SO(D-1,1)$
with integer (half odd integer) 
weights.} are 
quantized with respect to Bose (Fermi) statistics.

The CPT transformation actually comes in two varieties, one being  
the hermitean conjugate of the other: 

The first, which is what Pauli [\Ref{Pauli}] called 
{\it strong reflection} (SR), essentially maps a quantum field 
$\phi(t,\vec{x}) \rightarrow ({\rm phase}) \ \phi(-t,-\vec{x})$, 
i.e.\ it reverses time and space coordinates 
simultaneously. It also reverses the order of operators in an operator 
product and therefore cannot be represented by any 
operator acting on the particle states. 
It is a symmetry of the operator algebra only.

The second, which we will consider 
to be the CPT transformation proper, is obtained by performing first 
SR and then taking the hermitean conjugate.
The resulting operation clearly does not change the order of 
operators in an operator product but instead all ${\Bbb C}$-numbers 
are complex conjugated. It can be represented by an anti-unitary 
(i.e.\ unitary and anti-linear) operator $\Theta$ acting on the Hilbert 
space of physical particle states. 
It will map $\vert {\rm ``in"} \rangle$-states 
into $\vert {\rm ``out"} \rangle$-states. 

As is well known, the combination of the transformations
of charge conjugation, parity and time reversal (C~$+$~P~$+$~T), when 
defined, is equal to the combination of strong reflection and 
hermitean conjugation (SR~$+$~HC),\note{This is true provided that the 
phase for each transformation is chosen appropriately.} thereby 
justifying the name 
CPT for the latter transformation.

Following L\"{u}ders [\Ref{Luders}], we define the SR transformation  
for scalar, spinor and vector fields as follows
$$\eqalignno{ & \phi(x) \ \srarrow \ \phi(-x) & \nameali{sr} \cr
& \psi(x) \ \srarrow \ \varphi_{_{\rm SR}} \gamma^{D+1} \psi(-x) \qquad 
{\rm with} 
\qquad (\varphi_{_{\rm SR}})^2 = -(-1)^{D/2} \cr
& \overline{\psi}(x) \ \srarrow \ - \varphi_{_{\rm SR}}^* \overline{\psi}(-x) 
\gamma^{D+1} \cr
& \phi^{\mu} (x) \ \srarrow \ - \phi^{\mu} (-x) \ , \cr } $$
where $\gamma^{D+1}$ is the chirality matrix.
It is easy to verify that the free-field equations of motion and (anti-) 
commutation  relations are invariant under these transformations 
when we also define 
SR to {\it invert the order of the operators} and {\it 
leave ${\Bbb C}$-numbers unchanged}. At this point it is obviously
essential that the fields satisfy the spin-statistics
relation.

For complex fields the requirement that the free-field 
(anti-) commutation relations
are invariant under SR only fixes the form of the SR transformations 
\sr\ up to an overall phase factor; but for real fields the choice of 
$\varphi_{_{\rm SR}}$ is constrained by the requirement that the 
SR transformation should commute with hermitean conjugation and
only a sign ambiguity remains.~\note{For fermions the 
Majorana reality condition is only compatible with the massive Dirac
equation in dimensions $D=2+8k$ or $D=4+8k$, $k \in {\Bbb N}$.} 
The transformation laws given in 
\sr\ correspond to a choice of phase that satisfies this requirement for real 
fields. The dependence on the dimension that enters into the phase 
$\varphi_{_{\rm SR}}$ for the spinor field is due to the fact that the charge 
conjugation matrix commutes with $\gamma^{D+1}$ in dimensions
$D= 4k, k \in {\Bbb N}$ but anti-commutes in dimensions 
$D=2+4k, k\in {\Bbb N}$.

For the bosonic fields the sign is chosen to agree with what we 
obtain by applying the tensor transformation law to the transformation
$x \rightarrow -x$. Thus a field with $N$ vector indices would transform under 
SR with a phase $(-1)^N$. 

For a general vacuum expectation value of local field operators the 
statement of CPT invariance becomes
$$
\eqalignno{ & \langle 0 \vert\Phi_1 (x_1)  \ldots\Phi_N (x_N)\vert 0
\rangle& \nameali{cptinvariance} \cr
&\qquad\qquad\quad \eqcpt \ \langle 0 \vert  
\Theta \Phi_1 (x_1) \Theta^{-1} \ldots 
\Theta \Phi_N (x_N) \Theta^{-1}  \vert 0 \rangle^*  \cr
&\qquad\qquad\quad \ = \ \langle 0 \vert \left( \Phi_N (x_N) \right)^{\rm SR} 
\ldots \left( \Phi_1 (x_1) \right)^{\rm SR} \vert 0 
\rangle \cr &\qquad\qquad\quad 
\ = \ (-1)^{J_1} (-1)^{J_2} {\cal N} \langle 0 \vert \Phi_N (-x_N) 
\ldots \Phi_1 (-x_1) \vert 0 \rangle \ , \qquad\qquad \cr } 
$$
where $(\Phi_i (x_i))^{\rm SR}$ is defined by the free-field 
transformation laws \sr ,
where $J_1$ is the number of Lorentz vector indices, $J_2$ is the number 
of $\gamma^{D+1} = -1$ spinor indices, and
$$ 
{\cal N} = \left\{ \matrix{ i^{N_F} & {\rm if} & D = 4k, k \in 
{\Bbb N} \cr 
& & \cr  
(-1)^{N_{\downarrow}} & {\rm if} & D = 2+4k, k \in 
{\Bbb N} \ , \cr}   \right.  
\efr
with $N_F$ being the total number of fermions (which is even) and 
$N_{\downarrow}$ the number of covariant (lower) spinor indices ($\psi$ is
contravariant and $\overline{\psi}$ is covariant). 
The different behavior in $2$ and $4$ dimensions (mod $4$) 
is again due to the different chirality properties of the charge 
conjugation matrix. 

If the points $x_1,\ldots,x_N$ are such that all separations $x_i-x_k$, 
$i \neq k$, are space-like, and if we assume the validity of the 
spin-statistics theorem, the following condition, sometimes called {\it 
Weak Local Commutativity}, holds [\Ref{SW}]
$$ 
\langle 0 \vert \Phi_N (x_N) \ldots \Phi_1 (x_1) \vert 0 \rangle =
i^{N_F} \langle 0 \vert \Phi_1 (x_1) \ldots \Phi_N (x_N) \vert 0 \rangle 
\ . 
\nfr{wlc}
The phase appearing is just a sign, counting how many times two fermions 
have been transposed in the process of reversing the order of the 
operators. A priori this sign is $(-1)^{N_F(N_F-1)/2}$, but since $N_F$ is  
even, this equals $i^{N_F}$.

Combining eqs. \cptinvariance\ and \wlc\ we obtain another formulation 
of CPT invariance:
$$ \eqalignno{
& \langle 0 \vert \Phi_1 (x_1) \ldots \Phi_N (x_N) \vert 0 \rangle 
\ = & \nameali{cptandwlc} \cr & \qquad\qquad\qquad
i^{N_F} \langle 0 \vert \left( \Phi_1(x_1) \right)^{\rm SR} \ldots 
\left( \Phi_N (x_N) \right)^{\rm SR} \vert 0 \rangle \ , \cr} $$ 
valid when all separations are space-like.
\chapter{CPT Theorem
for two dimensional Field Theories on a genus zero surface.}
In this section we specialize to the case of two dimensions,
writing $x = (\tau,\sigma)$ and introducing light-cone coordinates
$\sigma^{\pm} = \tau \pm \sigma$. 
\section{CPT for a generic two-dimensional Field Theory.}
We begin by considering field theories in flat two-dimensional
Minkowski space-time that satisfy the standard CPT and spin-statistics
theorems outlined in section 1.

In two dimensions there is a very 
simple connection between the index structure of a field operator and 
the phase that it acquires under SR. Like in any (even) number of
dimensions, a field with $N$ vector 
indices (whether covariant or contravariant) picks up a phase $(-1)^N$
and spinors with negative and positive chirality transform with
opposite sign. Furthermore, since $\varphi_{_{\rm SR}} = \pm 1$ in two 
dimensions, it follows from eq.~\sr\
that a covariant and a contravariant spinor also 
transform with opposite sign (unlike in four dimensions where they transform 
with the same sign). Let us choose $\varphi_{_{\rm SR}}=1$ so that a
contravariant Weyl-spinor of positive chirality transforms
under SR with a plus sign.

Now, a general tensor field in $D=2$, with any combination of vector
and spinor indices, can be decomposed into component fields. Suppose we
decompose the vector indices into light-cone components $\sigma_+$ and
$\sigma_-$ and the Dirac spinors according to the chirality. Then, for
each component field we may define the quantities
$$ \Delta = N_+ - N^+ + {1 \over 2} (n^+ - n_+) \quad {\rm and} \quad
\overline{\Delta} = N_- - N^- + {1 \over 2} (n^- - n_-) \ , \efr
where $N_+$ and $N^+$ ($N_-$ and $N^-$) denote the number of covariant
and contravariant $\sigma_+$ ($\sigma_-$) indices, and likewise
$n_+$ and $n^+$ ($n_-$ and $n^-$) denote the number of covariant
and contravariant spinor indices of positive (negative) chirality.
Denoting the component field by $\Phi_{(\Delta,\overline{\Delta})}$ we
arrive at the very simple SR transformation law:
$$\eqalignno{ &
\Phi_{(\Delta,\overline{\Delta})} (\sigma^+,\sigma^-) \ \srarrow  &
\nameali{confsrone} \cr
& (-1)^{N^+ + N^- + N_+ + N_- + n^- + n_+} \ (-1)^{n_{_{\rm tot}}
(n_{_{\rm tot}}-1)/2} \ \Phi_{(\Delta,\overline{\Delta})} 
( -\sigma^+, -\sigma^- ) \ , \cr}$$
where $n_{_{\rm tot}} = n^+ + n_+ + n^- + n_-$ is the total number of
spinor indices. Here the first sign factor is obtained simply by 
putting a minus sign for each vector index, for each contravariant
spinor index of negative chirality and for each covariant spinor index
of positive chirality. The extra sign factor $(-1)^{n_{_{\rm tot}}
(n_{_{\rm tot}}-1)/2}$ is needed for fields with several spinor
indices. This can be seen if we imagine such a field to be given by a
normal-ordered product of $n_{_{\rm tot}}$ fermions, each having just
a single spinor index. SR inverts the order of these operators, and
the extra sign appears when they are permuted back into their original
order.

Eq.~\confsrone \ can be rewritten on the much simpler form
$$
\Phi_{(\Delta,\overline{\Delta})} (\sigma^+,\sigma^-) \ \srarrow \
(-1)^{\Delta -\overline{\Delta} - s}
\Phi_{(\Delta,\overline{\Delta})} ( -\sigma^+, -\sigma^- ) \ ,  
\nfr{confsr}
where $s=0$ or $s=1/2$ depending on whether $n_{_{\rm tot}}$ is even
($\Phi$ is bosonic) or odd ($\Phi$ is fermionic).

If we insert the transformation law \confsr \ into the CPT theorem
\cptinvariance \ and use the assumption \wlc \ of Weak Local
Commutativity to invert the order of the operators, we find that the
resulting identity \cptandwlc \ assumes the following suggestive form:
$$\eqalignno{ & \langle 0 \vert \Phi_{(\Delta_1,\overline{\Delta}_1)} 
(\sigma^+_1,\sigma^-_1) \ldots \Phi_{(\Delta_N,\overline{\Delta}_N)} 
(\sigma^+_N,\sigma^-_N) \vert 0 \rangle \ = &
\nameali{cptandwlctwodim} \cr
& (-1)^{\sum_{i=1}^N(\Delta_i - \overline{\Delta}_i)} \
\langle 0 \vert \Phi_{(\Delta_1,\overline{\Delta}_1)}
(-\sigma^+_1,-\sigma^-_1) \ldots 
\Phi_{(\Delta_N,\overline{\Delta}_N)} (-\sigma^+_N,-\sigma^-_N) \vert
0 \rangle \ . \cr } $$
Up to this point we have not assumed that the two-dimensional field theory 
is conformally invariant. For a CFT, the component field
$\Phi_{(\Delta,\overline{\Delta})}$ becomes a primary conformal field
of dimensions $(\Delta,\overline{\Delta})$, and
the phase factors in eq.~\cptandwlctwodim \ are seen to 
agree exactly with what one obtains from
applying the transformation law of a primary field to the 
transformation $(\sigma^+,\sigma^-) \rightarrow
(-\sigma^+,-\sigma^-)$. 

In summary, we have seen that the combined assumptions of CPT
invariance and Weak Local Commutativity lead to the identity 
\cptandwlctwodim \ that merely expresses the invariance under the 
conformal transformation $(\sigma^+,\sigma^-) \rightarrow 
(-\sigma^+,-\sigma^-)$.
Thus, for a two-dimensional field theory that is invariant under 
(complexified) conformal transformations, eq.~\cptandwlctwodim \ 
is always valid, regardless of whether Weak Local Commutativity holds 
and irrespective of the various assumptions underlying the standard CPT and
spin-statistics theorems.. Accordingly, inside the class of conformal field 
theories on the plane eq.~\cptandwlctwodim \ is a much more powerful
statement than the ordinary CPT theorem. 
The point of view which we will take,
and which will be elaborated further in the next subsection, is
therefore to abandon the assumptions of Lorentz invariance, (weak) local
commutativity and energy positivity, as well as the related assumption
of spin-statistics, and take instead as our one fundamental assumption
the invariance of the CFT under the transformation $\sigma_{\pm}
\rightarrow - \sigma_{\pm}$. In standard complex coordinates this
becomes the Belavin-Polyakov-Zamolodchikov (BPZ) transformation 
$z \rightarrow 1/z$~[\Ref{BPZ}], which will be studied in detail in
the following subsection. 

Abandoning Lorentz invariance has the great advantage of liberating
our discussion from the context of flat two-dimensional Minkowski
space-time. Instead, we can now study the genus zero
world-sheets of interest to string theory, such as the cylinder, where the 
space-direction has been compactified, and also surfaces 
with any number of external ``tubes'', corresponding to the presence of 
many incoming and outgoing strings. On all these surfaces global 
Lorentz invariance ceases to be meaningful, whereas the 
transformation $z \rightarrow 1/z$ remains well-defined. 

Since we abandon the requirement of spin-statistics on the world-sheet, 
we are also free to consider reparametrization ghosts and superghosts.
Finally, we may rotate to Euclidean world-sheet time, which is the
natural set-up for conformal field theory and hence for the
formulation of  world-sheet CPT 
based on the assumption of conformal invariance.

Having stressed the advantages it is only fair to point out what we
loose in the process. Since in eq.~\cptandwlctwodim \ the operators
appear in the same order on both sides of the equality sign,
the BPZ symmetry transformation $\sigma_{\pm} \rightarrow - 
\sigma_{\pm}$ must be defined {\it not} to invert the order of
operators. Thus, even though we may still define an anti-linear
world-sheet CPT transformation by combining the BPZ transformation 
$\sigma_{\pm} \rightarrow - \sigma_{\pm}$ with hermitean conjugation, 
the resulting transformation {\it will} invert the order of operators
and hence can never be generated by any operator acting on states. 
Thus, the world-sheet
CPT theorem for CFT is only an identity between correlation
functions. This is unlike the standard CPT theorem in Minkowski
space-time, where a major ingredient is the existence of an anti-unitary
operator implementing the CPT transformation on the Hilbert space of
particle states.
\section{BPZ invariance in conformal field theories.}
Consider a conformal field theory on a surface of genus zero. From 
the point of view of string theory, the interesting surfaces are
those with $N$ external ``tubes'' representing external string
states. These surfaces are conformally equivalent to the $N$-punctured sphere,
where the boundary conditions describing the external string
states are imposed by appropriate operators inserted at the punctures.
Therefore we will consider correlation functions of primary and 
descendant operator fields living on the sphere.

We introduce standard holomorphic coordinates $z$ and $\bar{z}$
related to $\sigma$ and $\tau$ by $z=\exp\{ i (\sigma+\tau)\}$ and
$\bar{z} = \exp\{i(-\sigma+\tau)\}$  
and rotate to Euclidean time $\tau \rightarrow -i \tau$.
We will refer to $z$ as a {\it global holomorphic coordinate} even
though, strictly speaking, $z$ is not defined at the point $z=\infty$,
where we use instead the coordinate $w=1/z$.
The map $\sigma^{\pm} \rightarrow - \sigma^{\pm}$ changes sign
on $\tau$ and $\sigma$ simultaneously and from a 
CFT point of view gives rise to the Belavin-Polyakov-Zamolodchikov (BPZ) 
transformation $z \rightarrow 1/z$ [\Ref{BPZ}]. This transformation defines a 
globally holomorphic diffeomorfism on the sphere. 

At the level of the operator fields, the transformation changes the 
coordinate system from $z$ to $w$ where $w=1/z$, and a primary conformal 
field $\Phi_{(\Delta,\overline{\Delta})}$ of conformal
dimension $(\Delta,\overline{\Delta})$ transforms as:
$$\eqalignno{ \Phi_{(\Delta,\overline{\Delta})} (z=\zeta,\bar{z}=\bar{\zeta}) 
\ \bpzarrow & \ \Phi_{(\Delta,\overline{\Delta})} 
(w=\zeta,\bar{w}=\bar{\zeta}) \ = & \nameali{bpz} \cr
& (-1)^{\Delta-\overline{\Delta}}
\left( {1 \over \zeta^2} \right)^{\Delta} 
\left( {1 \over \bar{\zeta}^2} \right)^{\overline{\Delta}} 
\Phi_{(\Delta,\overline{\Delta})} (z=1/\zeta,\bar{z}=1/\bar{\zeta}) \ .
\cr}$$
The BPZ transformation plays a role in CFT very similar to the role 
played by the SR transformation for field theories in flat Minkowski 
space-time: It maps an operator field located at the point
$(\tau,\sigma)$ into an operator field located at the point 
$(-\tau,-\sigma)$; it leaves all ${\Bbb C}$-numbers unchanged and 
cannot be represented by any operator acting on ket-states. 
But as we saw in subsection 2.1 there is one important difference:
Unlike the SR transformation, the BPZ transformation does 
{\it not\/} reverse the order of operators. 

Any conformal field theory on the sphere is invariant under the 
transformation $z \rightarrow w=1/z$. This implies the ``Ward 
identity''~\note{Here, and in most subsequent formulae, we restrict 
ourselves to chiral fields for ease of notation.}
$$
\langle \Phi_{\Delta_1} (z=\zeta_1)  
\ldots \Phi_{\Delta_N} (z=\zeta_N) \rangle \eqbpz
\langle \Phi_{\Delta_1} (w=\zeta_1)  
\ldots \Phi_{\Delta_N} (w=\zeta_N) \rangle   \ .
\nfr{bpzinvariance} 
This equation is the statement of BPZ invariance for a conformal field theory
on a genus zero Riemann surface and it holds for all fields, whether 
primary or descendant. For primary fields we may use eq.~\bpz \ to rewrite 
the identity \bpzinvariance \ as 
$$\eqalignno{&
\langle \Phi_{\Delta_1} (z=\zeta_1)  
\ldots \Phi_{\Delta_N} (z=\zeta_N) \rangle \eqbpz &\nameali{srtwod}\cr
&(-1)^{\Delta_1 + \ldots + \Delta_N}
\langle ({1\over\zeta_1})^{2\Delta_1} \Phi_{\Delta_1} 
(z={1 \over \zeta_1}) \ldots 
({1\over\zeta_N})^{2\Delta_N} \Phi_{\Delta_N} (z={1 \over \zeta_N}) 
\rangle  \ .\cr}
$$
When $\Delta - \overline{\Delta}$ is not integer, the field
$\Phi_{(\Delta,\overline{\Delta})}$ is said to be {\it non-local}.
For a non-local field the transformation \bpz\ is not well-defined a priori. 
To make it unambiguous we have to specify the phase of the complex
number $\zeta$ and we also have to choose a certain phase for $-1$.
Likewise, in a correlation function involving non-local fields, 
we need to specify the phases of the various differences
$\zeta_i - \zeta_j$. 
We would like to make these phase choices in such a way that
eqs.~\bpzinvariance \ and \srtwod \ remain valid. 
If we represent the explicit $-1$ appearing 
in the BPZ transformation \bpz \ by $e^{-i \epsilon \pi}$, so that the
overall factor in eq.~\srtwod \ becomes $e^{-i \epsilon \pi (\Delta_1
+ \ldots + \Delta_N)}$ (where $\epsilon$ is an odd integer), 
this is done by choosing
$$ {\zeta_j - \zeta_i \over \zeta_i - \zeta_j } = e^{-i \epsilon \pi}
\qquad {\rm for} \ {\it any} \ {\rm pair} \ 1 \leq i < j \leq N \ .
\nfr{condition}
In this sense, the BPZ
transformation and the statement of BPZ invariance can be extended to
non-local CFTs.

Notice that for integer and half odd integer conformal dimension,
the BPZ ``Ward identity'' \srtwod\ is nothing but 
eq.~\cptandwlctwodim\ transformed into $z$-coordinates.
Indeed, in the case of half odd integer conformal dimension
the choice of $\epsilon$ merely corresponds to the sign 
ambiguity also found in the choice of $\varphi_{_{\rm SR}}$ in eq.~\sr .
\section{Hermitean Conjugation and CPT on a genus zero surface.}
Just as we obtained CPT in the general $D$-dimensional case by composing 
SR with hermitean conjugation, so in the case of two-dimensional 
conformal field theories on the sphere we define what we call the WS-CPT 
transformation as the result of composing BPZ conjugation with hermitean 
conjugation (HC). But one should keep in mind that the WS-CPT
transformation thus defined will only be a formal substitution rule: 
Since it inverts the order of operators it can never be implemented by
any operator acting on states. 

To have a concept of hermitean conjugate 
it is necessary to assume that an inner product is defined on the 
Hilbert space of the CFT, 
i.e. for any two states $\vert \Phi_1 \rangle$ and $\vert \Phi_2 
\rangle$ we may form the complex number
$$ \langle \Phi_1 \vert \Phi_2 \rangle =
\langle \Phi_2 \vert \Phi_1 \rangle^* \ . \nfr{hermstate}
The basic hermiticity property \hermstate \ ensures that the norm of any 
state is a real number. We will not assume it to be positive. 

We can think of specifying the hermitean conjugate of all elementary 
operator fields in the conformal field theory by specifying the hermitean 
conjugate of the corresponding oscillators, with the further 
understanding that hermitean conjugation also complex conjugates all 
complex numbers and inverts the order of the operators.

For example, if
$$\Phi_{\Delta} (z=\zeta) = \sum_n \phi_n \zeta^{-n-\Delta} \efr
is a primary chiral conformal field of conformal dimension $\Delta$, then the 
hermitean conjugate of this field is
$$ \left( \Phi_{\Delta} (z=\zeta) \right)^{\dagger} \ = \
\left( {1 \over \zeta^*} \right)^{2\Delta} \widehat{\Phi}_{\Delta} 
(z=1/\zeta^*) \ , 
\nfr{eqhat}
where $\zeta^*$ denotes the complex conjugate of $\zeta$ (it is
sometimes convenient to think of $\zeta$ 
and $\bar{\zeta}$ as independent complex variables, so that $\zeta^*$ and 
$\bar{\zeta}$ need not be equal). The field $\widehat{\Phi}_{\Delta}$ is 
defined by
$$ \widehat\Phi_{\Delta} (z=\zeta) \equiv \sum_n \phi_{-n}^{\dagger} 
\zeta^{-n-\Delta} 
\nfr{hermmode}
and is called the hermitean conjugate of the field $\Phi_{\Delta}$.
We say that a field $\Phi_{\Delta}$ is hermitean (anti-hermitean) 
when $\widehat\Phi_{\Delta} = +\Phi_{\Delta} \ (-\Phi_{\Delta})$.
We always require the energy-momentum tensor of the CFT to be hermitean,
i.e. the mode operators 
satisfy $L_n^{\dagger} = L_{-n}$ for all $n \in {\Bbb Z}$. 
This in turn implies that $\widehat{\Phi}_{\Delta}$
is a primary conformal field of the same dimension as $\Phi_{\Delta}$.

The fact that HC changes the argument from $\zeta$ to $1/\zeta^*$  
means that an operator field situated in the vicinity of $z=0$ is mapped 
into one situated around $z=\infty$. For this reason, HC is only 
well-defined (as a map acting on the operator fields) in the case of the 
sphere, where we can think of $z$ as a {\it global} complex coordinate.
The peculiar behaviour of the argument  
is due to the fact that we are considering imaginary time on the 
world-sheet. Rotating back to real time, we have 
$z=\zeta=\exp\{ i(\tau +\sigma)\}$ and $1/\zeta^* = \zeta$.

In CFT we have the standard correspondence 
$$ 
\vert \Phi_{\Delta} \rangle \ \equiv \ \lim_{\zeta \rightarrow 0} 
\Phi_{\Delta} (z=\zeta) \vert 0 \rangle \ \efr
between states and the primary operator fields and their 
descendants. Therefore, if we know the vacuum expectation values (on 
the sphere) of arbitrary combinations of primary fields, and if we have 
defined the concept of hermitean conjugation of all operators, we may 
compute the norm of any two states as
$$ \langle \Phi_{\Delta_1} \vert \Phi_{\Delta_2} \rangle =
\lim_{\zeta_1 \rightarrow 0} \lim_{\zeta_2 \rightarrow 0}
\langle \left( \Phi_{\Delta_1} (z=\zeta_1) \right)^{\dagger}
\Phi_{\Delta_2} (z=\zeta_2) \rangle \ . \efr
{}From this point of view, the statement \hermstate\ of hermiticity 
becomes the requirement that 
$$
\langle \Phi_{\Delta_1}(z_1) \ldots \Phi_{\Delta_N}(z_{_N}) \rangle
\eqhc \left(\langle (\Phi_{\Delta_N}(z_{_N}))^\dagger \ldots 
(\Phi_{\Delta_1}(z_1))^\dagger \rangle\right)^*
\nfr{HC} 
for all correlation functions on the sphere. 

Combining eq.~\HC\ with eqs.~\bpzinvariance \ and 
\srtwod\ we obtain the form of the CPT theorem 
on a genus zero Riemann surface which we will also refer to as the  
World-Sheet CPT Theorem (WS-CPT) on the sphere
$$\eqalignno{ & \langle\, \Phi_{\Delta_1} (z=\zeta_1) \ldots 
\Phi_{\Delta_N} 
(z=\zeta_N) \,\rangle \ \eqwcpt \ \langle\,\left( \Phi_{\Delta_N} 
(w=\zeta_N) 
\right)^{\dagger} \ldots \left( \Phi_{\Delta_1} (w=\zeta_1) 
\right)^{\dagger} \,\rangle^*  \cr
& \qquad\qquad = (-1)^{\Delta_1 + \ldots + \Delta_N} 
\  \ \langle\,\widehat{\Phi}_{\Delta_N} (z=\zeta_N^*) \ldots 
\widehat{\Phi}_{\Delta_1} 
(z=\zeta_1^*) \,\rangle^* \ . & \nameali{wscpt} \cr} 
$$
Here the first equality sign holds for all fields whereas (a priori) the 
second equality sign holds only for primary fields. However, as can be 
seen by taking derivatives w.r.t. $\zeta_i$, $i=1,\ldots,N$, 
the second equality sign actually holds 
for descendant fields as well, as long as we let $\Delta_i$ denote the 
conformal dimension of the primary field from which $\Phi_{\Delta_i}$ 
descends. 

Like in the case of the BPZ identity \srtwod , 
eq.~\wscpt \ only holds for non-local fields if a proper
choice of phases is made: If we represent the explicit $-1$
appearing in the last line by $e^{- i \epsilon \pi}$ we have to impose
the condition \condition .

In the formulation \wscpt \ of the WS-CPT theorem the
operator $\Phi_{\Delta_i}$ is inserted at 
$z=\zeta_i$ whereas $\widehat{\Phi}_{\Delta_i}$ is inserted at 
$z=\zeta_i^*$. Thus,
unless $\zeta_i$ is real, the two operators are
inserted at {\it different} points on the sphere.
If we refer back to the real coordinates $(\tau,\sigma)$, in terms of
which
$$ z(\tau,\sigma) = e^{\tau + i \sigma} = \zeta \ , \nfr{ztausigma}
this point of view corresponds to saying that $\zeta$ is mapped into
$\zeta^*$ on account of $(\tau,\sigma)
\rightarrow (\tau,-\sigma)$.

We see that there is another, equivalent, way of looking at it: Namely
to say that the geometrical point $(\tau,\sigma)$ remains fixed, but
the complex structure is changed by the
hermitean conjugation so that the holomorphic coordinate is
now given by
$$ \tilde{z} (\tau,\sigma) = e^{\tau - i \sigma} = (z(\tau,\sigma))^* \ ,
\nfr{ztildets}
instead of eq.~\ztausigma .

The two points of view differ only by the diffeomorfism $(\tau,\sigma)
\rightarrow (\tau,-\sigma)$ which in terms of the coordinate $z$ maps
$z=\zeta$ into $z=\zeta^*$. Diffeomorfism invariance 
of the correlation functions ensures that
$$\eqalignno{ & \langle\,\Phi_{\Delta_1} (\tilde{z}=\zeta_1) \ldots 
\Phi_{\Delta_N} (\tilde{z}=\zeta_N)\,\rangle^{(\tilde{M})} \ = 
& \nameali{diffeoinv} \cr
&\qquad\qquad\qquad
\langle\, \Phi_{\Delta_1} (z=\zeta_1) \ldots 
\Phi_{\Delta_N} (z=\zeta_N)\,\rangle^{(M)} \ , \cr}$$
where the labels $(M)$ and $(\tilde{M})$ are introduced to remind us
that the two correlators pertain to different complex structures.
Eq.~\diffeoinv \ is really just the statement that the 
correlation function depends only on the {\it value} taken at 
the operator insertion points by the global holomorphic coordinate, and 
not on how this coordinate is related to some underlying 
real coordinates. 

Using eq.~\diffeoinv , the statement \wscpt \ can be formulated
as follows
$$\eqalignno{ & \langle\,\Phi_{\Delta_1} (z=\zeta_1) \ldots \Phi_{\Delta_N} 
(z=\zeta_N)\,\rangle^{(M)} \ \eqwcpt & \nameali{wscpttwo} \cr
&\qquad\qquad e^{-i\epsilon \pi (\Delta_1 + \ldots + \Delta_N)}\ \left(
\langle\,
\widehat{\Phi}_{\Delta_N} (\tilde{z}=\zeta_N^*) \ldots 
\widehat{\Phi}_{\Delta_1} (\tilde{z}=\zeta_1^*)\,\rangle^{(\tilde{M})}
\right)^* \ , \cr}$$
where now $z=\zeta$ and $\tilde{z}=\zeta^*$ refer to the same point
$(\tau,\sigma)$ on the sphere. We have explicitly parametrized $-1$ by
the odd integer $\epsilon$ in accordance with the phase choice
\condition . Thus, eq.~\wscpttwo \ also holds for non-local theories.

In the following section we consider how to extend the formulation
\wscpttwo \ of WS-CPT invariance to higher genus surfaces.
\chapter{The World-Sheet CPT Theorem on a Genus $g$ Surface.}
In the case of Riemann surfaces of genus $g > 0$ the transformation 
$z \rightarrow w=1/z$ is not anymore a global symmetry, i.e. does not 
define any globally holomorphic diffeomorfism. Thus the concept of BPZ
invariance looses its meaning.~\note{The case $g=1$ is 
special. 
If we represent the torus of modular parameter $k=\exp \{ 2\pi i \tau 
\}$ by an annulus in the complex 
plane, $|k| \leq |z| \leq 1$, then the modular transformation changing 
sign on the two homology cycles can be represented by the
globally holomorphic diffeomorfism $z \rightarrow 1/z$. In this way
the BPZ transformation can be extended to the torus.}
Nor can we introduce a
globally well-defined (i.e. single-valued) ho\-lo\-mor\-phic coordinate $z$,
like the one employed in the previous section. This means that our
definition \eqhat \ of the hermitean conjugate of a field operator is
not readily generalized either.

Nevertheless, as we shall see in this section,
the world-sheet CPT theorem, when written on the form
$$\eqalignno{ & \langle\,\Phi_{\Delta_1} (z_1=0) \ldots \Phi_{\Delta_N} 
(z_N=0)\,\rangle^{(M)}_g \ \eqwcpt & \nameali{wscptlocalg} \cr
&\qquad\qquad (-1)^{\Delta_1 + \ldots + \Delta_N}\ \left(
\langle\,
\widehat{\Phi}_{\Delta_N} (\tilde{z}_N=0) \ldots 
\widehat{\Phi}_{\Delta_1} (\tilde{z}_1=0)\,\rangle^{(\tilde{M})}_g 
\right)^* \ , \cr}$$
remains valid for local, modular-invariant CFTs satisfying a very mild
assumption (given by eq.~(3.7) below). 
The correlator on the left-hand side of eq.~\wscptlocalg \ refers to a 
Riemann surface $M$ endowed with {\sl local\/} holomorphic
coordinates $z_i$, while that on 
the right-hand side 
refers to a Riemann surface $\tilde{M}$ endowed with {\sl local\/} 
holomorphic coordinates $\tilde{z}_i$. 

The {\sl local\/} coordinates $z_i$ on $M$ depend holomorphically on each 
other (whenever the regions on which they are defined overlap). In terms of
some fixed holomorphic coordinate $z$ we may write 
$z_i = z_i (z,z^*) = z_i (z)$. 
Then the {\sl local\/} coordinate $\tilde{z}_i$ defined on $\tilde{M}$
is given by the following anti-holomorphic function of $z$
$$ \tilde{z}_i (z^*) = (z_i (z))^* \ . \efr
The relation between the two sets of coordinates is more easily
expressed by referring back to some fixed set of real coordinates,
$\xi$. In terms of these
$$ \tilde{z}_i (\xi) = (z_i (\xi))^* \ . \nfr{zitildetwo}
Thus, the point of view underlying the formulation \wscptlocalg \
of WS-CPT invariance is that $M$ and $\tilde{M}$ correspond to the
same real two-dimensional manifold ${\cal M}$, but endowed with
different complex structures $J$ and $\tilde{J}$, and
$z_i=\zeta$ and $\tilde{z}_i=\zeta^*$
correspond to the same geometrical point on ${\cal M}$
(the same value of the coordinate $\xi$).

In the case of the sphere any two complex structures are related by a
diffeomorfism and in this sense $M$ and $\tilde{M}$ are therefore
equivalent. At genus $g \geq 1$ this will in general
{\it not} be the case, $M$ and $\tilde{M}$
will correspond to different points in moduli space.

The Riemann surface $\tilde{M}$ is referred to as the {\it mirror-image 
of} $M$. A general discussion of mirror image Riemann surfaces 
can be found in the Appendix.

The world-sheet CPT theorem \wscptlocalg\ naturally leads us to define
an anti-linear 
two-dimensional (world-sheet) CPT transformation for the primary and 
descendant operator fields of a CFT by
$$
\Phi_{\Delta} (z_i=0) \ \wcptarrow \ 
\left( \Phi_{\Delta} (z_i=0) \right)^{\rm WS-CPT} \ 
\equiv \  (-1)^{\Delta}
\widehat{\Phi}_{\Delta} (\tilde{z}_i=0) \ , \nfr{cpttrans}
but one should keep in mind that 
this is a formal substitution rule, making sense only inside a
correlation function, rather than a genuine CPT 
transformation: First of all, it relates operator fields defined on
different Riemann surfaces ($M$ and $\tilde{M}$ respectively). Second,
it involves an inversion of the ordering of operators, so  
even when $M$ and $\tilde{M}$ are equivalent (as in the case of the sphere)
it can never be represented by any operator acting on states. 

Assuming the validity of the WS-CPT theorem on the sphere there is a 
straightforward way to extend it also to higher genus Riemann surfaces.
The way to do this is just by ``{\sl sewing\/}'' Riemann surfaces, using 
the sewing procedure for conformal field theories defined in 
refs.~[\Ref{Sonodatwo}]. 
If we start from a genus $g$ surface with $N+2$ punctures 
and sew together two of the legs, we obtain a genus
$g+1$ surface with $N$ punctures, and any $N$-point correlator on this 
genus $g+1$ surface can be expressed as a sum over $N+2$-point 
correlators on the genus $g$ surface.

Thus, assuming that the WS-CPT theorem is valid on the sphere, we may 
use induction to prove that it is valid at any genus. All we have to do 
is to show that, assuming the WS-CPT theorem to hold for the correlators 
at genus $g$, then it also holds for the genus $g+1$ correlators defined 
from these by sewing.

We will carry out this analysis in two cases: First 
for a large class of local conformal field theories (to be specified 
more precisely in the following subsection), 
then for some important 
examples of free non-local theories. Finally, we consider some 
convention-dependent complications that arise for the combined theory of 
left- and right-moving reparametrization ghosts.
\section{Local Conformal Field Theories}
In this section we adopt Sonoda's formulation of sewing [\Ref{Sonodatwo}]
for local conformal field theories, that is, CFTs where primary fields 
have integer dimensions (or $\Delta -\bar\Delta\in {\Bbb Z}$ for non-chiral
CFT). We always have in mind a CFT describing some consistent string
model. We exclude the reparametrization ghosts, which need special treatment
and will be discussed in subsection 3.4. The resulting CFT is always local:
The vertex operators describing on-shell physical states are 
primary fields with $\Delta=\overline{\Delta}=1$, and even if we go 
off-shell, only states satisfying the level-matching condition 
$L_0 - \bar{L}_0 = 0$ can propagate.

Consider a Riemann surface $M'$ with genus $g$ and $N$ punctures. Add 
two extra punctures, $P$ and $Q$, to this surface with local coordinates
$z'$ and $w'$ vanishing at $P$ and $Q$ respectively. Let $q$ be a complex
parameter with $\vert q\vert < 1$ (the {\it sewing parameter}). 
The standard procedure is to 
excise two discs around  
the two punctures and identify the coordinates as $z'w'=q$, obtaining 
in this way a genus $g+1$ surface, $M$. We obtain the correlators on the 
genus $g+1$ surface from the similar correlators on the genus $g$ 
surface by inserting at the points $P$ and $Q$ a complete set of 
conjugate local operators:
$$\eqalignno{&
\vev{\Phi_{\Delta_1}(z_1=0) \ldots \Phi_{\Delta_N}(z_N=0)}_{g+1}^{(M)} \ = 
&\nameali{Sonsewing}\cr  
&\sum_{m,n} \left({\cal M}^{-1}_{wz}\right)_{nm}
\vev{\Phi_{\Delta_1}(z_1=0) \ldots
\Phi_{\Delta_N}(z_N=0) \Phi_n (z'=0) \Phi_m(w'=0)}_{g}^{(M')} \ , \cr}
$$
where by definition
$$ ({\cal M}_{wz})_{nm} = \vev{ \Phi_n (w=0) \Phi_m (z=0) }_{g=0}
\ \ \ , \ \ \
({\cal M}_{zw})_{nm} = \vev{ \Phi_n (z=0) \Phi_m (w=0) }_{g=0} \ , \efr
the coordinates $z$ and $w$ being related by $w=1/z$. The two matrices
differ by the phase factor $(-1)^{2\Delta_n}$. For the local CFTs
considered in the present subsection this phase is always unity,
whereas for the non-local theories considered in the next subsection
it will be non-trivial. 

In eq.~\Sonsewing \ the $\{ \Phi_n \}$ form a basis for the set of 
primary fields and their descendants. In practical applications, this set 
is often specified by means of an appropriate projection defined inside a 
larger class of fields. For example, if we imagine the CFT to describe a 
superstring theory in the Neveu-Schwarz Ramond formulation, then only 
the GSO projected fields are included in the sum.
The sewing formula
\Sonsewing \ is manifestly independent of the choice of basis.

A genus $g+1$ surface
has 3 more moduli than a genus $g$ surface. These are the parameter $q$
and the positions of the two points $P$ and $Q$, more precisely, the
values taken at these points by an appropriate holomorphic
coordinate.~\note{The two well-known exceptions are $g=0$ and $g=1$
where the existence of global conformal diffeomorfisms makes it possible
to fix the coordinate values of $P$ and $Q$ (for $g=0$) and either $P$
or $Q$ (for $g=1$).}

Notice that on the right-hand side of the sewing formula \Sonsewing \
the local operator inserted at $P$ appears to the left of the local
operator inserted at $Q$. This relative ordering of the two operators
distinguishes between the points $P$ and $Q$ and 
can be used to define an orientation
of the homology cycle $b=b_{g+1}$ formed by the sewing. Therefore, it
would be more precise to denote the surface obtained from the sewing
procedure \Sonsewing \ as $M^P_Q$. The surface $M_P^Q$ can be obtained
from $M_Q^P$ by interchanging the points $P$ and $Q$ on $M'$, i.e. by
reversing the orientation of the homology cycle $b$ on $M$. This
corresponds to a modular transformation, and $M^P_Q$ and $M^Q_P$ 
therefore define the same point in moduli space but different points
in Teichm\"{u}ller space. 

If we only assume locality, there is no guarantee that the correlation
functions defined by the sewing formula \Sonsewing \ will be modular
invariant. Also, the correlation functions on a given Riemann surface
will in general depend on the precise way in which the surface is
sewn together from three-punctured spheres. 
Sonoda~[\Ref{Sonodatwo}] proved that both of these problems are
avoided if one further assumes i) crossing invariance and ii)
that the one-point genus one correlators are modular invariant. 

Crossing invariance implies that correlation functions do not depend on the
order of the operator fields. From the point of view of string theory this
requirement is somewhat too restrictive, since space-time fermions are
described by vertex operators which satisfy Fermi statistics on the
world-sheet~[\Ref{uscpt}]. For this reason we will not assume crossing
invariance. 

Instead, we take the point of view that the correlation
functions on the sphere are known and that the higher-loop
correlation functions are then defined recursively by the explicit sewing
formula \Sonsewing . {\it We restrict ourselves to CFTs for which} 
eq.~\Sonsewing \ {\it leads to modular invariant correlation
functions}. Obviously, the CFTs 
considered by Sonoda fall into this category. So does any CFT
describing a consistent string theory. Such a CFT is also expected to
satisfy duality, i.e. 
the requirement that correlation functions obtained on higher
genus surfaces do not depend on the details of the sewing, except
for overall numerical factors.

The absence of crossing invariance means that we have to keep track of
the ordering of the operators. In particular, the precise location of the 
two ``internal'' operators $\Phi_n$ and $\Phi_m$ inside the 
correlator in eq.~\Sonsewing \ is important. 
The locations chosen are in accordance with the
sewing procedure of refs.~[\Ref{PDV1},\Ref{PDV2}] and with eq.~(2.15)
of ref.~[\Ref{Sonoda}]. 

For the purpose of proving the WS-CPT theorem \wscptlocalg \ for the 
correlation functions defined by eq.~\Sonsewing \ 
we need one more technical assumption,
which is much weaker than crossing invariance: 
{\it For any pair of operators} $\Phi_n$ {\it and} $\Phi_m$
{\it such that} $({\cal M}^{-1}_{wz})_{mn}$ {\it is nonzero and for any set of
operators} $\Phi_{\Delta_1} \ldots \Phi_{\Delta_N}$ {\it having a nonzero
correlator at genus} $g+1$ {\it we require that}
$$ \eqalignno{ & \vev{\Phi_{\Delta_1}(z_1=0) \ldots
\Phi_{\Delta_N}(z_N=0) \Phi_n (z'=0) \Phi_m(w'=0)}_{g} \ = &
\nameali{assumption} \cr
& \qquad \qquad \vev{\Phi_n (z'=0) \Phi_m(w'=0) \Phi_{\Delta_1}(z_1=0) \ldots
\Phi_{\Delta_N}(z_N=0) }_{g} \ . \cr }$$
We are now ready to prove the WS-CPT theorem \wscptlocalg \ at any genus.
Assume it to be valid at genera $\leq g$.
Then we may rewrite the right-hand side of eq.~\Sonsewing \ to obtain
$$\eqalignno{ & \vev{ \Phi_{\Delta_1} (z_1=0) \ldots \Phi_{\Delta_N}
(z_N=0) }_{g+1}^{(M^P_Q)} \ = & \nameali{RHSone} \cr
&\sum_{m,n} \left({\cal M}^{-1}_{wz}\right)_{nm}
\vev{ \Phi_{\Delta_1}(z_1=0) \ldots
\Phi_{\Delta_N}(z_N=0) \Phi_n (z'=0) \Phi_m(w'=0)}_{g}^{(M')} \ = \cr
& \sum_{m,n} \left( {\cal M}^{-1}_{wz} \right)_{nm} \left( \vev{ 
 \tilde{\Phi}_{\Delta_N} (\tilde{z}_N =0) \ldots
\tilde{\Phi}_{\Delta_1} (\tilde{z}_1=0) \tilde{\Phi}_m
(\tilde{w}'=0) \tilde{\Phi}_n (\tilde{z}'=0)
}_g^{(\tilde{M'})}  \right)^* \ , \cr} $$
where the assumption \assumption , as well as the
inductive hypothesis, was used in the last line. We defined
$\tilde{\Phi}_{\Delta} = (-1)^{\Delta} \widehat{\Phi}_{\Delta}$.
The $\{ \tilde{\Phi}_n \}$ also form a
basis for the set of primary fields and their descendants, 
and in terms of this basis
we have a matrix $(\widetilde{\cal M}_{wz})_{mn}$.

WS-CPT invariance \wscptlocalg \ and diffeomorfism invariance \diffeoinv \
ensure that
$$\eqalignno{({\cal M}_{wz})_{nm} = &\vev{\Phi_n (w=0) \Phi_m (z=0) } \ = \
\vev{\tilde{\Phi}_m (\tilde{z}=0) \tilde{\Phi}_n (\tilde{w}=0) }^* \ =
& \nameali{helpid} \cr
& \vev{\tilde{\Phi}_m (z=0) \tilde{\Phi}_n (w=0) }^* \ =
\ (\widetilde{\cal M}_{zw})_{mn}^* \ . \cr }$$
Therefore, we may rewrite eq.~\RHSone \ in the form
$$\eqalignno{ & \vev{\Phi_{\Delta_1} (z_1=0) \ldots \Phi_{\Delta_N}
(z_N=0) }_{g+1}^{(M^P_Q)}  \ = \ & \nameali{threeseven} \cr
& \left( \sum_{m,n} \left( \widetilde{M}^{-1}_{zw} \right)_{mn}
\vev{ \tilde{\Phi}_{\Delta_N}
(\tilde{z}_N=0) \ldots \tilde{\Phi}_{\Delta_1} (\tilde{z}_1=0)
\tilde{\Phi}_m (\tilde{w}'=0) 
\tilde{\Phi}_n (\tilde{z}'=0)}_g^{(\tilde{M}')}  \right)^* \ = \cr
& \left( \vev{\tilde{\Phi}_{\Delta_N} (\tilde{z}_N=0) \ldots
\tilde{\Phi}_{\Delta_1} (\tilde{z}_1=0)}_{g+1} ^{(\tilde{M}^Q_P)} \right)^* \ .
\cr}$$
Here we used the sewing formula \Sonsewing \ ``backwards'', keeping in
mind that it is irrelevant which basis we choose for the set
of local operators. 

When we compare the surface $M^P_Q$ defined from $M'$ through 
eq.~\Sonsewing \ with the surface $\tilde{M}^Q_P$ 
obtained from $\tilde{M}'$ by means of eq.~\threeseven , we first of
all notice that $P$ and $Q$ have been interchanged, implying 
a change of sign in the homology cycle $b_{g+1}$.  This will be
unimportant only if the correlation functions are invariant under the
corresponding modular transformation. As already mentioned, we
explicitly assume the correlation functions defined by eq.~\Sonsewing
\ to be modular invariant.

We further notice that 
all local coordinates on $\tilde{M}$ are related to those on $M$ by the 
standard map \zitildetwo . Thus, $\tilde{M}$ is indeed the mirror
image of $M$. In particular, since $z'w'=q$, we have 
$\tilde{z}' \tilde{w}' = q^*$, so that $\tilde{M}$ has sewing parameter 
$q^*$ rather than $q$. 
This concludes the proof of the WS-CPT theorem for higher genus Riemann
surfaces.
\section{Free Non-local Conformal Field Theories}
The analysis in the previous section was restricted to local conformal field
theories, meaning that $\Delta-\overline{\Delta} \in {\Bbb Z}$ for all
primary fields. 
If we relax these conditions, things get much more complicated:
The operator fields become multivalued and to properly describe this 
we have to introduce spin structures. 

We do not attempt to analyze this problem in general, instead we
consider some examples of free, chiral conformal field theories that are
important building blocks for constructing superstring theories in the
Neveu-Schwarz Ramond formulation: A pair of real world-sheet fermions or
a single complex world-sheet fermion; equivalently, a boson with
appropriately discretized momenta; and the
superghosts $(\beta,\gamma)$. These CFTs are all non-local, but since
they are free, they can be treated very explicitly.

We denote the spin-structure around the $\mu$'th $a$-cycle ($b$-cycle) by
$\alpha_{\mu}$ ($\beta_{\mu}$), $\mu=1,\ldots,g$.
Our conventions are that $\alpha=0$
denotes a Ramond sector and $\alpha=1/2$ denotes a Neveu-Schwarz sector.
For the superghosts, $\alpha$ and $\beta$ can only take the values $0$
and $1/2$ (mod $1$), but for a free complex fermion other possibilities are
allowed.

The correlation functions at genus $g$ depend on the spin
structure. When we go from genus $g$ to genus $g+1$ by means of
sewing, an extra pair of components $\alpha = \alpha_{g+1}$ and
$\beta=\beta_{g+1}$ are added to the vectors ${\bfmath \alpha}$ and
${\bfmath \beta}$. If we denote the spin structures of the genus $g$
surface $M'$ by ${\bfmath \alpha}'$ and ${\bfmath \beta}'$, and those
of the genus $g+1$ surface $M$ by ${\bfmath \alpha}$ and ${\bfmath
\beta}$, the sewing formula \Sonsewing \ is modified as follows
$$\eqalignno{&
\vev{\Phi_{\Delta_1}(z_1=0) \ldots \Phi_{\Delta_N}
(z_N=0)}_{g+1}^{(M^{P}_{Q})}
\left[{}^{{\bfmath\alpha}}_{{\bfmath\beta}}\right]
\ = &\nameali{Sonsewtre} \cr  & 
{\sum_{m,n} }^{(\alpha)}\left( {\cal M}^{-1}_{wz} \right)_{n m} \
\vev{ \ \Phi_{\Delta_1} (z_1=0) \ldots \Phi_{\Delta_N}
(z_N=0) \Phi_{m} (w'=0) \ P_{\beta} \
\Phi_{n} (z'=0) }_{g}^{(M')} 
\left[{}^{{\bfmath\alpha}'}_{{\bfmath\beta}'}\right]\ . \cr}
$$
This is essentially the sewing formula of ref.~[\Ref{PDV1}] rephrased in
the language of Sonoda. Notice that it differs from the sewing formula
\Sonsewing \ of the previous subsection in various ways.

First of all, the operators $\Phi_n (z'=0)$ and $\Phi_m (w'=0)$ appear
in the opposite relative order. Since we consider non-local theories,
the two orderings differ by a non-trivial phase. The ordering 
in eq.~\Sonsewtre \ has been carefully chosen to reproduce the known
formulae for the correlation functions. 

The $\{ \Phi_{n} \}$ in eq.~\Sonsewtre \
form a basis for the set of all primary and 
descendant fields existing {\it in the sector labelled by the
spin structure} $\alpha$ (mod $1$) (this is indicated by the label
${}^{(\alpha)}$ appearing in the summation symbol), 
and the dependence on the spin structure
$\beta$ enters through the operator~[\Ref{PDV1}]
$$ P_{\beta} \equiv \exp \{ 2\pi i ({1 \over 2} + \beta) 
J_0 \} \ , \nfr{pbeta}
where $J_0$ is the number operator. The values of $J_0$ carried by the
operators $\Phi_n$ are as follows, 
$$ J_0 \in {\Bbb Z} + {1 \over 2} - \alpha \ . \nfr{eigenvalues}
Whether we consider a pair of real fermions, a single complex fermion or
the superghost system, we may bosonize, introducing a free boson field
$\phi$; the number operator is then given by
$$J_0 = \oint_0 {{\rm d}z \over 2\pi i} \partial \phi (z) \ . \nfr{jzerodef}
We want to investigate to which extent the WS-CPT identity
\wscptlocalg , valid for local modular-invariant CFTs satisfying
the assumption \assumption , carries over to the higher genus correlation
functions defined by eq.~\Sonsewtre .

We start by noticing that the assumption \assumption \ is satisfied
for the theories under consideration in this subsection.  
In the bosonized formulation this follows from the fact that 
both the product $\Phi_n (w'=0) \Phi_m (z'=0)$, as well as the product
of all ``external'' operators, $\Phi_{\Delta_1} (z_1=0) \ldots
\Phi_{\Delta_N} (z_N=0)$, carry vanishing total value of $J_0$ (mod
$2$).
This is the statement of fermion number/superghost number/momentum
conservation, depending on which theory we are considering.

For the same reason $P_{\beta}$ commutes with $\Phi_{\Delta_1} (z_1=0) \ldots
\Phi_{\Delta_N} (z_N=0)$.
Even so, the presence of $P_{\beta}$ 
in the sewing formula \Sonsewtre \ influences
the behaviour of the higher genus correlation functions under WS-CPT. 
To see exactly how, we first have to study how $J_0$ transforms under
the WS-CPT transformation \cpttrans . Using the
definitions \jzerodef \ and \hermmode \ we obtain
$$ J_0 = \oint {{\rm d} \zeta \over 2\pi i} \partial \phi (z=\zeta) \
\wcptarrow \ \oint {{\rm d} \zeta^* \over 2\pi i} (-1)
\widehat{\partial \phi} (z=\zeta^*) = - J_0^{\dagger} \ . \efr
The operator $J_0$ is hermitean for ``ordinary'' fermions, i.e. for 
fermions whose mode 
operators in the Neveu-Schwarz sector give rise to a Fock space with 
positive definite norm. 
The fermions describing the transverse (and possibly internal) degrees of 
freedom of a superstring are of this type.

Instead, $J_0$ is anti-hermitean ($J_0^{\dagger} = - J_0$)
for the pair of real fermions,
$\psi^0$ and $\psi^1$, that are related (by world-sheet supersymmetry) 
to the time-like and longitudinal space-time coordinate fields $X^0$
and $X^1$ 
(in Minkowski space-time) [\Ref{mink}].

Finally, in the superghost case $J_0^{\dagger} = -J_0 +2$, 
i.e. the number operator is anti-hermitean mod $2$.

In summary, the number operator $J_0$ has the following behaviour under 
WS-CPT:
$$\eqalignno{ & J_0 \wcptarrow - J_0 \qquad {\rm \ for \ an \ 
ordinary \ fermion \ theory,} & \numali \cr
& J_0 \wcptarrow J_0 \qquad {\rm \ \ \ \ \ 
for \ the \ fermion \ pair} \ \psi^0 \ {\rm and} \ \psi^1 \ ,  
\cr
& J_0 \wcptarrow J_0 - 2 \qquad {\rm for \ the \ superghosts.} \cr}$$
Thus, there are basically two cases to consider: In the first case $J_0$ is 
hermitean, which implies that $P_{\beta}$ is invariant under the
anti-linear 
WS-CPT transformation; in the second case $J_0$ is anti-hermitean (mod
2), meaning that
$$ P_{\beta} = e^{2\pi i (1/2+\beta) J_0} \wcptarrow e^{-2\pi i 
(1/2+\beta) J_0} =  (-1)^{(1+2\beta)(1-2\alpha)} 
\ P_{\beta}\ . \efr
Here we used the property \eigenvalues \ of the $J_0$ values carried
by $\Phi_{m}$ (and 
the fact that $2\alpha, 2\beta \in {\Bbb Z}$).

The two cases also differ from each other in one more respect:
If $J_0$ is hermitean, it changes sign under WS-CPT. By eq.~\eigenvalues 
\ this implies that WS-CPT maps the operators in the sector labeled 
by $\alpha$ into the sector labeled by $-\alpha$, i.e.
if $\{ \Phi_{n} \}$ is a basis of the primary and descendant fields 
with spin structure $\alpha$, then $\{ \tilde{\Phi}_{n}
\}$ is a basis of the 
primary and descendant fields 
with spin structure $-\alpha$ (mod $1$). 
Whereas if $J_0$ is anti-hermitean (mod 2), both
$\Phi_{n}$ and $\tilde{\Phi}_{n}$ will have the same 
spin structure $\alpha$. 

Finally we notice that, since the correlation functions pertaining to
individual spin structures are not modular invariant, 
the distinction between the points $P$
($z'=0$) and $Q$ ($w'=0$) 
in the sewing formula \Sonsewtre \ is important.  

In view of our various considerations we may now formulate the
identities that replace the WS-CPT theorem \wscptlocalg \ at genus $g$
for the free non-local CFTs under consideration:

Let
$M^{\bfmath P}_{\bfmath Q}$, where ${\bfmath P} = \{ P_{\mu} \vert
\mu=1,\ldots,g\}$ and ${\bfmath Q} = \{ Q_{\mu} \vert
\mu=1,\ldots,g\}$, denote a surface obtained from the
sphere by $g$ successive sewings. Then, if $J_0$ is hermitean, 
we have the identity
$$\eqalignno{ & \langle \, \Phi_{\Delta_1} (z_1=0) \ldots
\Phi_{\Delta_N} (z_N = 0) \, \rangle_{g}^{(M^{\bfmath P}_{\bfmath Q})}
\left[{}^{\bfmath\alpha}_{\bfmath\beta}\right]
 \ \eqwcpt
&\nameali{wscptgf}\cr
&e^{-i\epsilon \pi (\Delta_1 + \ldots + \Delta_N)} \
\left(\langle \, \widehat{\Phi}_{\Delta_N} (\tilde{z}_N = 0)
\ldots \widehat{\Phi}_{\Delta_1} (\tilde{z}_1 =0)
\, \rangle_{g}^{(\tilde{M}^{\bfmath Q}_{\bfmath P})}
\left[{}^{-\bfmath\alpha}_{\ \bfmath\beta}\right] \right)^*\ . \cr}
$$
Instead, for the pair of fermions describing the time- and the
longitudinal space-direction, and for the superghosts, we have the identity
$$\eqalignno{ & \langle \, \Phi_{\Delta_1} (z_1=0) \ldots
\Phi_{\Delta_N} (z_N = 0) \, \rangle_{g}^{(M^{\bfmath P}_{\bfmath Q})}
\left[{}^{\bfmath\alpha}_{\bfmath\beta}\right]
 \ \eqwcpt
&\nameali{wscptgftwo}\cr
&(-1)^{P({\bfmath \alpha},{\bfmath \beta})} \
e^{-i \epsilon \pi (\Delta_1 + \ldots + \Delta_N)}
\ \left(\langle \, \widehat{\Phi}_{\Delta_N} (\tilde{z}_N = 0)
\ldots \widehat{\Phi}_{\Delta_1} (\tilde{z}_1 =0)
\, \rangle_{g}^{(\tilde{M}^{\bfmath Q}_{\bfmath P})}
\left[{}^{\bfmath\alpha}_{\bfmath\beta}\right] \right)^*\ , \cr}
$$
where $P({\bfmath \alpha},{\bfmath \beta}) = \sum_{\mu=1}^g
(1-2\alpha_{\mu})(1+2\beta_{\mu})$ so that the sign $(-1)^{P({\bfmath
\alpha},{\bfmath \beta})}$ is $+1$ ($-1$) for even (odd) spin
structures.

Both identities can be formally proven by induction, following a
similar line of arguments as in the previous subsection, but starting now
from the sewing formula \Sonsewtre . 

Like in the case of the identity \wscpttwo \ on the sphere, 
the identities \wscptgf \ and
\wscptgftwo \ are only valid if an appropriate $\epsilon$-dependent
phase choice is made, generalizing eq.~\condition . In terms of the
local holomorphic coordinates vanishing at the punctures the correct
requirement is that
$$ {E(z_j=0,z_i=0) \over E(z_i=0,z_j=0) } = e^{-i \epsilon \pi}
\quad {\rm for} \ {\it any} \ {\rm pair} \ 1 \leq i < j \leq N \ .
\nfr{condition}
Here $E$ denotes the prime form on $M$, which has the short distance
behaviour $E(z=\zeta_1,z=\zeta_2) = 
\zeta_1-\zeta_2 + {\cal O} (\zeta_1-\zeta_2)^2$.

When we combine our various non-local CFTs into a consistent string
theory and sum over the spin structures with appropriate weights, 
locality and modular invariance should be restored, 
and hence WS-CPT invariance in the 
sense of eq.~\wscptlocalg . 

In the light-cone gauge, we do not
have the time-like/longitudinal fermion pair, nor the superghosts.
Thus any free complex fermion (or free boson) involved will
satisfy eq.~\wscptgf . For example, we may consider the heterotic string
models of refs.~[\Ref{KLT},\Ref{Anto},\Ref{Bluhm}] 
where (apart from the space-time coordinate fields) 
all degrees of freedom are described by a set of 
free complex fermions, $\{ \psi \}$. These models were constructed to
be modular invariant and the spin structure summation coefficients 
$C^{\{ \bfmath \alpha \} }_{\{ \bfmath \beta \} }$ multiplying the
correlation functions obtained from the sewing formula \Sonsewtre \ are
known explicitly. 
They satisfy 
$$
\left( C^{\{ \bfmath \alpha \} }_{\{ \bfmath \beta \} } \right)^*
= C^{\{ -{ \bfmath \alpha} \} }_{\{ \bfmath \beta \} } \ , \efr
which is seen to be exactly the property needed in order for the 
string correlation functions  to obey eq.~\wscptlocalg\ after having
performed the sum over the spin-structures.~\note{When a given
correlation function of the KLT model is decomposed into a product
of correlation functions for the constituent free fermion theories, it
is of course essential to include the appropriate cocycle factors. 
We explicitly
checked that these factors do not affect the validity of the WS-CPT theorem.}

It can also be verified that 
when the sewing formula \Sonsewtre \ is applied to the
entire KLT model and the sum over the spin structures is performed,
one recovers the sewing formula \Sonsewing .

If we now consider the Lorentz-covariant formulation, by
adding the two CFTs describing, respectively, the superghosts, and
the time-like and longitudinal
components of $\psi^{\mu}$, we see that the WS-CPT invariance of the 
string correlation functions are not 
affected:  Both theories have the ``anomalous'' behaviour \wscptgftwo \ ,
i.e.\ involve an extra minus sign for odd spin structures. But this extra 
sign always cancels when we consider the product of the correlation 
functions pertaining to the two theories.

\section{An Explicit Example: Free boson or fermion}
For the free CFTs considered in the previous subsection explicit 
formulae are known for all correlation functions at genus $g$. Therefore 
we may verify the identities \wscptgf \ and \wscptgftwo \ simply 
by inserting the known expressions for the correlation functions 
involved. In this subsection we perform such a check. 
We consider a pair of components of the world-sheet fermion
$\psi^{\mu}$ (which is the world-sheet super-partner of the space-time
coordinate field $X^{\mu}$). We bosonize this pair in terms of a free
boson $\phi$, which is anti-hermitean if the fermion pair corresponds to
two transverse components (the case $J_0^{\dagger} = J_0$)
and hermitean in the case of the fermion pair
$\{ \psi^0, \psi^1 \}$ (where $J_0^{\dagger} = - J_0$).

The correlation function we consider is~[\Ref{PDV1}]
$$\eqalignno{
& \langle\, \prod_{l=1}^N e^{a_l\phi (\zeta_l)}
\,\rangle_{g}^{(M^{\bfmath P}_{\bfmath Q})} 
\left[{}^{\bfmath\alpha}_{\bfmath\beta}\right]  \ =\ & 
\nameali{corrfunc} \cr
& \quad 
\delta_{\sum_l a_l,0} \ ({\det}' \bar{\partial}_0)^{-1/2} 
\ \prod_{1 \leq j<k \leq N}(E(\zeta_j,\zeta_k))^{a_ja_k} \
\Teta{\bfmath\alpha}{\bfmath\beta}
\left(\sum_{l=1}^N a_l \int^{\zeta_l} \frac{\omega}{2\pi i}\vert \tau \right)
\ . \cr}$$
The various quantities appearing in this formula can be written
down explicitly if we use the Schottky parametrization, see
ref.~[\Ref{PDV1}] for details.~\note{Our conventions for
spin structures and $\Theta$ functions differ from those of
ref.~[\Ref{PDV1}] and can be found in ref.~[\Ref{normaliz}].}
In the Schottky parametrization, a higher genus
Riemann surface $M$ is represented by the sphere, $S^2$, endowed with the
usual global holomorphic coordinate $z$, modulo a discrete
symmetry group, the Schottky group ${\cal G}(M)$~[\Ref{PDV1}], 
generated by $g$ projective
transformations $S_{\mu} (z)$, each of which can be parametrized by
the multiplier $k_{\mu}$ together with the fixed
points $\eta_{\mu}$ and $\xi_{\mu}$.~\note{More precisely, the points
on $M$ are in one-to-one correspondence with $S^2$ {\it minus}
the limit set of the Schottky group, modulo the action of 
${\cal G}(M)$~[\Ref{Bers}].} 

Specifying the generators $S_{\mu}$ (up to an overall projective
transformation of the coordinate $z$) defines not only a Riemann surface
but also a canonical homology basis. The modular transformation
$P_{\mu} \leftrightarrow Q_{\mu}$ changing sign on all the $2g$ cycles
$(a_{\mu},b_{\mu})$ in a canonical homology basis corresponds to
the interchange $\eta_{\mu} \leftrightarrow \xi_{\mu}$, which replaces
each generator $S_{\mu} (z)$ by the inverse map, $S_{\mu}^{-1} (z)$.
Thus, if we associate the surface $M^P_Q$ with the set of generators
$\{ S_{\mu} \}$, the surface $M^Q_P$ can be associated with the set of
generators $\{ S_{\mu}^{-1} \}$.

WS-CPT maps the holomorphic coordinate $z$ into $\tilde{z}=z^*$ and therefore
relates the Riemann surface $M$ to the ``mirror image''
Riemann surface $\tilde{M}$, whose Schottky group ${\cal
G}(\tilde{M})$ is generated by the
transformations $\tilde{S}_{\mu} (\tilde{z}) = \tilde{S}_{\mu} (z^*)
\equiv (S_{\mu} (z))^*$ having multipliers $k_{\mu}^*$ and 
fixed points $\eta_{\mu}^*$ and $\xi_{\mu}^*$. Accordingly, we may
identify
$$ (\tilde{M}^{\bfmath Q}_{\bfmath P}) \sim \{ \tilde{S}_{\mu}^{-1} \}
\quad {\rm and}
\quad  (\tilde{M}^{\bfmath P}_{\bfmath Q}) \sim \{ \tilde{S}_{\mu} \} \ . \efr
Starting from eq.~\corrfunc\ (and assuming for simplicity 
that $2\alpha,2\beta\in{\Bbb Z}$)
it is straightforward to verify the following two identities, by using 
the explicit expressions for all relevant quantities given in 
refs.~[\Ref{normaliz},\Ref{PDV1}]:
$$
\langle \prod_{l=1}^N e^{a_l\phi (\zeta_l)}
\rangle_{g}^{\{S_{\mu}\}}\left[{}^{\bfmath\alpha}_{\bfmath\beta}\right] \ = \
(-1)^{P({\bfmath \alpha},{\bfmath \beta})}\
\left(\langle \prod_{l=N}^1
e^{+i\pi\epsilon\Delta_l}e^{a_l\hat\phi (\zeta_l^*)}
\rangle_{g}^{\{\tilde{S}_{\mu}\}}\left[{}^{\bfmath\alpha}_{\bfmath\beta}
\right]\right)^* \ ,
\nfr{checkone}
if $\phi$ is anti-hermitean; and
$$\langle \prod_{l=1}^N e^{a_l\phi (z_l)}
\rangle_{g}^{\{S_{\mu}\}} \left[{}^{\bfmath\alpha}_{\bfmath\beta}\right] \ = \
\left(\langle \prod_{l=N}^1
e^{+i\pi\epsilon\Delta_l} e^{a_l\hat\phi (z_l^*)}
\rangle_{g}^{\{\tilde{S}_{\mu}\}}\left[{}^{\bfmath\alpha}_{\bfmath\beta}
\right] \right)^* \ ,
\nfr{checktwo}
if $\phi$ is hermitean.

The overall signs encountered on the right-hand sides of these two 
identities are opposite to those appearing in eqs.~\wscptgf \ and 
\wscptgftwo . The reason for this is that in eqs.~\checkone \ and 
\checktwo \ we compare $M^{\bfmath P}_{\bfmath Q}$ with
$\tilde{M}^{\bfmath P}_{\bfmath Q}$, whereas in eqs.~\wscptgf \ and 
\wscptgftwo \ we compared $M^{\bfmath P}_{\bfmath Q}$ with
$\tilde{M}^{\bfmath Q}_{\bfmath P}$.

Under the modular transformation $S_{\mu} \rightarrow S_{\mu}^{-1}$ 
that changes sign on all canonical homology cycles, 
the period matrix $\tau$ and the prime form $E$ remain unchanged, whereas the 
holomorphic one-forms $\omega_{\mu}$ change sign, and hence so does
the argument of the $\Theta$ function in eq.~\corrfunc . Since the 
$\Theta$ function 
is even (odd) for even (odd) spin structures, this produces the extra 
factor $(-1)^{P({\bfmath \alpha},{\bfmath \beta})}$ needed to recover 
eqs.~\wscptgf \ and \wscptgftwo , where the ``anomalous'' sign for odd 
spin structures appeared in the case where $\phi$ was hermitean.

In a similar way one can investigate what happens for the superghosts.
In this case the formula \corrfunc \ is modified (see 
ref.~[\Ref{Verlinde}]) but it can be verified that eq.~\checktwo \
remains valid and that the modular transformation $S_{\mu} \rightarrow
S_{\mu}^{-1}$ still gives rise to a minus sign for odd spin
structures. Thus we recover eq.~\wscptgftwo \ also in this case.
\section{WS-CPT for the Reparametrization Ghosts}
The reparametrization ghosts constitute an important ingredient of the
full CFT underlying the Lorentz-covariant formulation of any string
theory. For closed string theories we have to consider the combined
$(b,c)$ and $(\bar{b},\bar{c})$-system and in this case there is a
small complication
regarding the hermiticity property \HC , 
one of the two key ingredients of WS-CPT invariance.

The problem arises when considering the basic non-vanishing correlator
on the sphere,
$\vev{\bar{c}_{-1}\bar{c}_{0}\bar{c}_{1}c_{-1}c_{0}c_{1}}_{g=0}$.
Since $c_n^{\dagger} = c_{-n}$ and $\bar{c}_n^{\dagger} = \bar{c}_{-n}$ the
operator involved is explicitly anti-hermitean. 
Thus, in order for the correlation
functions to satisfy the hermiticity property \HC \ it is necessary
to postulate that this correlator is imaginary, for example
$$
\vev{\bar{c}_{-1}\bar{c}_{0}\bar{c}_{1}c_{-1}c_{0}c_{1}}_{g=0} \ =\ i\ .
\nfr{convone}
With this convention the discussion of subsections (2.3) and (3.1) goes
through without modification and the standard formulation
\wscptlocalg \ of WS-CPT invariance holds.

The convention \convone \ is rather inconvenient in practical calculations.
Instead, one can adopt a different convention where
the basic non-vanishing correlator is real,
$$
\vev{\bar{c}_{-1}\bar{c}_{0}\bar{c}_{1}c_{-1}c_{0}c_{1}}_{g=0} \ =\ 1\ .
\nfr{convtwo}
The price one has to pay is that an unusual minus sign appears in the
hermiticity formula \HC , in the statements \wscpt \ and \wscpttwo \
of WS-CPT invariance on the sphere, and hence also in the identity
\helpid . This implies that whenever we add a loop
by means of the sewing formula \Sonsewing \
an extra minus sign enters into the statement of WS-CPT invariance.

In summary, the convention \convtwo \ leads to $(b,c,\bar{b},\bar{c})$
correlation functions satisfying
$$\eqalignno{ &
\langle \Phi_{\Delta_1} (z_1=0) \ldots \Phi_{\Delta_N} (z_N=0)
\rangle_g^{(M)} \ \eqwcpt & \nameali{ghostwscpt} \cr
& \qquad (-1)^{g+1} \ (-1)^{\Delta_1+\ldots+\Delta_N} \
\left( \langle \widehat{\Phi}_{\Delta_N}
(\tilde{z}_N=0) \ldots \widehat{\Phi}_{\Delta_1}  (\tilde{z}_1 = 0)
\rangle_g^{(\tilde{M})} \right)^* \ . \cr}$$
\chapter{WS-CPT in String Theory and Space-Time Hermiticity.}
\setchap{hermsection}
In this section we consider a generic string theory in $D$-dimensional
Minkowski space-time, based on a CFT which is assumed to satisfy the WS-CPT
theorem \wscptlocalg \ (or eq.~\ghostwscpt \ if the reparametrization
ghosts satisfy eq.~\convtwo ),
and we study the physical meaning of the WS-CPT transformation
\cpttrans \ when
applied to the vertex operator of a physical string state. We also show
that WS-CPT invariance of the underlying CFT implies that the scattering
$T$-matrix is formally hermitean.

We define the $T$-matrix element as the connected $S$-matrix element
with certain normalization factors
removed
$$
\eqalignno{
& { \langle \rho_1, \dots ; in \vert S \vert \dots , \rho_{_N} ; in
\rangle_{\rm connected} \over
\prod_{i=1}^N \left( \langle \rho_i ; in \vert
\rho_i ; in \rangle \right)^{1/2} }\  = & \nameali{Smatrix} \cr
&\qquad i (2\pi)^D \delta^D ( p_1+ \dots - p_N ) \ \prod_{i=1}^N
(2 p_i^0 V)^{-1/2} \  T ( \rho_1; \dots \vert\dots ; \rho_{_N}) \ , \cr }
$$
where $N$ is the total number of
external states, $p_i$ is the momentum of the $i$'th string state, all
of them having $p_i^0 > 0$, and $V$ is the
usual volume-of-the-world factor.  
The Minkowski metric is $\eta = {\rm diag}(-1,+1,\ldots ,+1)$.

 The $g$-loop contribution to the
$T$-matrix element is given by the Polyakov path
integral. If, for the sake of being definite, we consider a heterotic
superstring model, this is equivalent to the following operator
formula~\note{the details of which can be found in ref. [\Ref{normaliz}].}
$$
\eqalignno{ & T^g (\rho_1; \dots  \vert\dots;\rho_{_N} ) \ =
& \nameali{Tmatrix} \cr
&\qquad  (-1)^{g-1} C_g \int_{\cal D} \left(\prod_{I=1}^{3g-3+N}
\di^2 m^I \right) \
\vev{\left| \prod_{I=1}^{3g-3+N} (\eta_{m^I} \vert b)
\prod_{i=1}^N c(z_i=0) \right|^2\ \times \cr
&\qquad \left(\prod_{A=1}^{N_{_{\rm PCO}}}
\Pi (z_A^{^{\rm PCO}}=0)\right) \, {\cal V}_{\langle \rho_1 \vert }^{(q_1)}
(z_1=0)
\dots {\cal V}_{\vert \rho_N \rangle}^{(q_N)} (z_N=0) }^M_g  \ . \cr}
$$
Here the integral is over the moduli space of $N$-punctured genus $g$
Riemann surfaces $M$. The $\{m^I\}$, $I=1,\ldots, 3g-3+N$, 
is a set of holomorphic coordinates on moduli space,
defined on a domain ${\cal D} \in {\Bbb C}^{3g-3+N}$,
and $\eta_{m^I}$ is the Beltrami
differential corresponding to $m^I$. We may think of $m^I$ as a
global holomorphic coordinate, and of ${\cal D}$ as a fundamental domain
of the modular group, defined inside Teichm\"{u}ller space (see also
the Appendix).

By definition the correlator $\vev{\dots}$ in eq.~\Tmatrix \
includes the partition function
and the summation over spin structures.

To each incoming string state
$\vert \rho \rangle$ (each outgoing string state $\langle \rho \vert$)
is associated a BRST-invariant vertex operator
${\cal W}_{\vert\rho\rangle}^{(q)} = c \bar{c} {\cal
V}_{\vert\rho\rangle}^{(q)}$
(${\cal W}_{\langle\rho\vert }^{(q)} = c \bar{c} {\cal
V}_{\langle\rho\vert}^{(q)}$) of conformal dimension $\Delta =
\overline{\Delta} = 0$. Here $(q)$ denotes the picture (the superghost
charge), and the number $N_{_{\rm PCO}}$ of Picture Changing Operators
(PCOs) $\Pi$ is given by the integer
$$ N_{_{\rm PCO}} = 2g-2 - \sum_{i=1}^{N} q_i \ . \nfr{npco}
Since the PCOs have conformal dimension zero they do not depend on the 
choice of coordinate system. In eq.~\Tmatrix \ they are evaluated in 
terms of some arbitrary local holomorphic coordinates $z_A^{^{\rm PCO}}$.
We assume that the PCO insertion points do not depend on the moduli
(meaning that they have fixed values in a moduli-independent 
coordinate system).

The vertex operators are evaluated in terms of local holomorphic 
coordinates $z_i$, $i=1,\ldots,N$. 
In ref. [\Ref{normaliz}] it was
shown that the vertex operator ${\cal W}_{\langle \rho \vert}^{(q)}$ is
related to ${\cal W}_{\vert \rho \rangle}^{(q)}$ by
$$
{\cal W}_{\langle\rho\vert }^{(q)}(z_i=0)\ =\
(-1)^{q+1} \widehat{\cal W}_{\vert\rho\rangle}^{(q)}(z_i=0) \ ,
\nfr{inout}
where for pictures of half odd integer $q$ (i.e. pictures describing space-time
fermions) the phase factor $(-1)^{q+1}$ involves the choice of a sign which
will not matter in what follows.

Since the on-shell vertex operators are dimension $\Delta=\bar\Delta=0$
operators, we can interpret eq. \inout\ as
$$
{\cal W}_{\langle\rho\vert }^{(q)}(\tilde{z}_i=0) \ = \
(-1)^{q+1} \left({\cal W}_{\vert\rho\rangle}^{(q)}(z_i=0)
\right)^{\rm WS-CPT} \ .
\nfr{wsinout}
Recalling that $c \bar{c}$ is anti-hermitean this is equivalent to
$$
{\cal V}_{\langle\rho\vert }^{(q)}(\tilde{z}_i = 0) \ = \
(-1)^{q} \left({\cal V}_{\vert\rho\rangle}^{(q)}(z_i=0)
\right)^{\rm WS-CPT} \nfr{intertwo}
or
$$
{\cal V}_{\vert\rho\rangle}^{(q)}(\tilde{z}_i=0) \ =\
(-1)^q \left({\cal V}_{\langle\rho\vert}^{(q)}(z_i =0)
\right)^{\rm WS-CPT} \ .
\nfr{intertre}
Equation \wsinout\ gives the string theory interpretation of the WS-CPT
transformation: Up to a picture dependent phase factor
WS-CPT maps the vertex operator that describes a given incoming string
state into the vertex operator that we use in the scattering formula
\Tmatrix \ to 
describe the same string state when it is outgoing. This
interpretation is consistent
with the fact that WS-CPT maps ${\cal W}_{\vert \rho \rangle}^{(q)}$
into $\widehat{\cal W}_{\vert \rho \rangle}^{(q)}$:
If the vertex
operator ${\cal W}_{\vert \rho \rangle}^{(q)}$ creates a state
$\ket{\rho}=\ket{p,\eta,\{\lambda\}}$, where $p$ is the momentum, $\eta$
the ``helicity''~\note{In general, if the momentum points in the 
$(D-1)$-direction, we may think of
$\eta$ as the eigenvalues of the Lorentz generators $M^{12}, M^{34}, \ldots ,
M^{D-3,D-2}$.}
and $\{\lambda\}$ a set of gauge and enumerative quantum
numbers, then (as described in ref. [\Ref{uscpt}]) the 
operator $\widehat{\cal W}_{\vert \rho \rangle}^{(q)}$, when
acting on the conformal vacuum $\vert 0 \rangle$, creates the state
$\ket{-p,-\eta,\{-\lambda\}}$ (up to a phase).
This is what one expects from the fact that the new operator should describe
an outgoing string state with the same quantum numbers as $\vert \rho
\rangle$.

We now turn to consider the consequences of the WS-CPT theorem on the
string scattering amplitudes. In ref.~[\Ref{normaliz}] it was noticed
that eq.~\inout\ gives the correct hermiticity properties for the $S$-matrix
elements at tree level. Indeed,
unitarity requires that the tree-level $T$-matrix element is hermitean except
when the momentum flowing in some intermediate channel happens to be on
the mass-shell corresponding to some physical state in the theory.
In field theory the imaginary part appears as a result of the
$i\epsilon$-prescription present in the propagator that happens to be
on-shell. In string theory the tree amplitudes are expressed by an
integral over the Koba-Nielsen variables that will in general diverge but
which can be treated by an appropriate analytic continuation of the  
invariant energy variables. The analytic continuation can be chosen so 
that the resulting amplitude has the correct physical poles. 

The point of view of this paper is that the hermiticity of the tree-level
$T$-matrix elements away from the resonances is a direct
consequence of the WS-CPT theorem, as emphasized by Sonoda~[\Ref{Sonoda}].

Indeed, we can now show that
the WS-CPT theorem formulated in section~3
implies that the $T$-matrix elements, as given by eq.~\Tmatrix ,
are {\it formally hermitean\/} at
any loop order, more precisely, that
$$
\left(T^g(\rho_N; \dots \vert \dots;\rho_1)\right)^*\ =\
T^g(\rho_1; \dots \vert \dots;\rho_N)\ .
\nfr{eqstar}
The reason why the proof is only formal is that the integral over the
moduli in eq.~\Tmatrix \ is not always convergent. This is very fortunate,
since otherwise we would be in contradiction with unitarity.
The divergent contributions to the modular integral offers a way out:
When properly regularized, they should give rise to an imaginary part
of the $T$-matrix, as required by $S$-matrix unitarity. Explicit
regularizations accomplishing this in various cases, mainly at
one-loop level, have been 
proposed~[\Ref{Amano},\Ref{Mitchell},\Ref{Hoker},\Ref{Berera},\Ref{Weisberger}].

To make the formal proof of eq.~\eqstar , we start from eq.~\Tmatrix ,
according to which
$$\eqalignno{
&\left(T^g(\rho_N; \dots \vert \dots;\rho_1)\right)^*\ =\ &\numali\cr
&(-1)^{g-1} C_g^* \int_{{\cal D}^*} \left(\prod_{I=1}^{3g-3+N}
\di^2 m^{*I} \right) \
\left(\vev{\left| \prod_{I=1}^{3g-3+N} (\eta_{m^I} \vert b)
\prod_{i=1}^{N} c(z_i=0) \right|^2\ \times\right. \cr
&\qquad \left.\left(\prod_{A=1}^{N_{_{\rm PCO}}}
\Pi (z_A^{^{\rm PCO}}=0)\right) \, {\cal V}_{\langle \rho_N \vert }^{(q_N)} 
(z_N=0)
\dots {\cal V}_{\vert \rho_1\rangle}^{(q_1)} (z_1=0)}_g^{(M)}
\right)^* \ .\cr}
$$
We can now use the WS-CPT theorem to obtain
$$\eqalignno{
&\left(T^g(\rho_N; \dots \vert \dots;\rho_1)\right)^*\ =\
(-1)^{g+1} (-1)^{g-1} \, C_g  \  \times & \nameali{expression} \cr
& \int_{{\cal D}^*} \left(\prod_{I=1}^{3g-3+N}
\di^2 m^{*I} \right) \ \vev{ ({\cal V}_{\vert \rho_1 \rangle }^{(q_1)}
(z_1=0))^{\rm WS-CPT} \dots ({\cal V}_{\langle \rho_N\vert}^{(q_N)}
(z_N=0))^{\rm WS-CPT}\ \times \cr
&\left(\prod^{1}_{A=N_{_{\rm PCO}}} 
(\Pi (z_A^{^{\rm PCO}}=0))^{\rm WS-CPT}\right)
\left(\left| \prod_{I=1}^{3g-3+N} (\eta_{m^I} \vert b)
\prod_{i=1}^{N} c(z_i=0) \right|^2
\right)^{\rm WS-CPT}}_g^{(\tilde{M})}\, .
\cr}$$
Here we assumed that the normalization constant $C_g$ is real. This is
only true if we use the convention \convtwo \ for the ghost 
correlators~[\Ref{normaliz}].
Therefore we also had to include
the ``anomalous'' sign $(-1)^{g+1}$ that appears in the
WS-CPT identity \ghostwscpt \ for reparametrization ghosts. Since the
WS-CPT transformation inverts the order of the ghost operators, this sign is
cancelled when we put the ghost operators back into
their original order. The PCOs are bosonic operators and can be
rearranged without introducing any signs.
If we further use eqs.~\intertwo \ and \intertre ,
together with the fact that
$$\left(\Pi(z_A^{^{\rm PCO}}=0)\right)^{\rm WS-CPT}\ 
=\ -\Pi(\tilde{z}_A^{^{\rm PCO}}=0)  \ , \efr
we may rewrite eq.~\expression \ as
$$\eqalignno{
&\left(T^g(\rho_N; \dots \vert \dots;\rho_1)\right)^*\ =\
(-1)^{\sum_i q_i} (-1)^{N_{_{\rm PCO}}} \times
& \nameali{expressiontwo} \cr
&(-1)^{g-1} C_g \int_{{\cal D}^*} \left(\prod_{I=1}^{3g-3+N}
\di^2 m^{*I} \right) \ 
\vev{ {\cal V}_{\langle \rho_1 \vert }^{(q_1)}
(\tilde{z}_1=0) \dots {\cal V}_{\vert \rho_N\rangle}^{(q_N)}
(\tilde{z}_N=0) \ \times\cr
&\left(\prod_{A=1}^{N_{_{\rm PCO}}} \Pi (\tilde{z}_A^{^{\rm PCO}}=0)\right)
\left| \prod_{I=1}^{3g-3+N} (\eta_{m^{*I}} \vert b)
\prod_{i=1}^{N} c(\tilde{z}_i=0) \right|^2 }_g^{(\tilde{M})}\ .\cr}
$$
Here $(-1)^{N_{_{\rm PCO}} + \sum_i q_i} = +1$ by eq.~\npco ; and the
PCO and reparametrization ghost factors may be moved around, so as to appear
in the same places as in eq.~\Tmatrix , without introducing any extra
signs. 

By definition, the $\tilde{z}_i$ and $\tilde{z}_A^{^{\rm PCO}}$ are 
local holomorphic coordinates pertaining to the Riemann surface $\tilde{M}$;
As explained in greater detail in the appendix, a coordinate 
system $\{ m \}$ on moduli space, which assigns  
a Riemann surface $M$ to each set of moduli $\{ m \}$, automatically 
defines another coordinate system on moduli space, where the mirror 
image Riemann surface $\tilde{M}$ is assigned to the complex 
conjugate set of moduli, $\{ m^* \}$;  and 
the complex conjugate of the Beltrami differential 
$\eta_{m^I}$ on $M$ is exactly the Beltrami differential $\eta_{m^{*I}}$ 
on $\tilde{M}$. In summary, eq.~\expressiontwo \ is nothing but the
expression \Tmatrix \ for $T^g(\rho_1; \dots \vert \dots;\rho_N)$,
merely written in terms of the coordinates $\{ m^* \}$, defined on the
domain ${\cal D}^*$, rather than the $\{ m \}$, defined on the 
domain ${\cal D}$. 

Since eq.~\Tmatrix \ does not depend on which set of 
holomorphic coordinates we use to describe $N$-punctured genus $g$ 
moduli space, this concludes our proof that the $T$-matrix is
formally real at any loop order. Of course, the integral over
moduli space is not always convergent, and then the above proof breaks
down. Indeed, we see that in order to recover unitarity
it is {\it essential} that the integral over the moduli diverges
in the kinematical regions where the $T$-matrix is required to develop
an imaginary part. To handle this divergence, a regularization of the
integral is needed, and it is exactly the failure of the regularization
to be invariant under complex conjugation that restores the $S$-matrix
unitarity.
\appendix{Mirror Image Riemann Surfaces}
\setchap{Appconv}
In this appendix we give some details concerning mirror image Riemann 
surfaces.
\section{Moduli Space Generalities}
A Riemann surface $M$ can be defined as a compact, real, oriented 
two-dimensional manifold, ${\cal M}$, endowed with a complex structure 
$J$. We may write $M=({\cal M},J)$ for short. 
A complex structure is a tensor field of rank $(1,1)$, i.e. in terms of some 
real coordinate system $\xi^r$, $r=1,2$, it has real components 
$J_r^{\ s} (\xi)$. These are required to satisfy
$$ J_r^{\ s'} (\xi) J_{s'}^{\ s} (\xi) = - \delta_r^{\ s} \ . \efr
Given a complex structure we may define complex, holomorphic coordinates
as the solutions to Beltrami's equation
$$ J_r^{\ s} (\xi) {\partial z \over \partial \xi^s} = i {\partial 
z \over \partial \xi^r} \ . \nfr{Beltrami}
Without specifying any boundary conditions, 
eq.~\Beltrami \ does not determine the holomorphic coordinate $z$ 
uniquely. But it is rather easy to verify that if $z_1 (\xi)$ and $z_2 
(\xi)$ are two solutions, then one depends holomorphically on 
the other, i.e.
$$ {\partial z_1 \over \partial {z}_2^*} = 0 \ . \efr
Thus, Beltrami's equation specifies the holomorphic coordinate up to 
conformal coordinate transformations.

It is easy to see that the complex structure $J$, when evaluated in any 
coordinate system $z$ solving Beltrami's equation, is simply given by
the conformally invariant expressions
$$ J_z^{\ z} = i \qquad J_{\bar{z}}^{\ \bar{z}} = -i \qquad J_z^{\ 
\bar{z}} = J_{\bar{z}}^{\ z} = 0 \ . \nfr{jzz}
Moduli space is the set of complex structures modulo diffeomorfisms. That 
is, two complex structures $I$ and $J$ are considered equivalent from 
the point of view of moduli space if and only if there 
exists a diffeomorfism $\Phi$: ${\cal M} \rightarrow {\cal M}$ such that
(in terms of the coordinate system $\xi$)
$$ {\partial \Phi^{r'} (\xi) \over \partial \xi^r} \, I_{r'}^{\ s} (\Phi 
(\xi)) =  J_r^{\ s'} 
(\xi) \, {\partial \Phi^s (\xi) \over \partial \xi^{s'}}
\ , \nfr{diffequiv}
or, in a convenient shorthand notation, $I \congphi J$.

As always, a diffeomorfism is required to be differentiable and 
invertible. If we consider compact Riemann surfaces 
with $N$ punctures, i.e. with marked points $P_i$, $i=1,\ldots,N$, then the
diffeomorfisms $\Phi$ are also required to keep these points fixed.
 
The inequivalent complex structures on a genus $g$ surface with $N$
punctures span a moduli space of real 
dimension $6g-6+2N$. 
Local coordinates on moduli space can be introduced by assigning (in a 
differentiable way) a complex structure $J_m$ to each point 
$m=m^a$ belonging to some open subset of  ${\Bbb R}^{6g-6+2N}$,
$$ m^a \rightarrow J_r^{\ s} (m^a;\xi) = (J_m)_r^{\ s} (\xi) \ , \efr
in such a way that $J_m$ and $J_{m'}$ are inequivalent unless $m=m'$. 

If the complex structure $J$ appearing in Beltrami's equation \Beltrami 
\ depends on $m$, then so will the complex coordinate $z=z_m$ solving 
this equation. By differentiating eq.~\Beltrami \ with respect to $m^a$ 
(keeping $\xi$ fixed) we find the relation
$$ -2i \left( \eta_{m^a} \right)_r^{\ s} (m;\xi) \, {\partial 
z_m (\xi) \over \partial \xi^s}  + J_r^{\ s} (m;\xi) {\partial \over \partial 
\xi^s} {\partial z_m (\xi) \over \partial m^a}  - i {\partial \over 
\partial \xi^r} {\partial z_m (\xi) \over \partial m^a } = 0  
\nfr{bdifquasi}
between the so-called quasiconformal vector field
$\partial z_m (\xi) / \partial m^a$, measuring the change in the 
conformal coordinate, and the Beltrami differential
$$ \left( \eta_{m^a} \right)_r^{\ s} (m;\xi) \ \equiv \ 
{-1 \over 2i} {\partial \over 
\partial m^a} J_r^{\ s} (m;\xi) \nfr{beltramidif}
that parametrizes the change in the conformal structure. 
If we evaluate eq.~\bdifquasi \ in the holomorphic coordinates $z=z_m$, it 
reduces to
$$ \left( \eta_{m^a} \right)_{\bar{z}}^{\ z} (z,\bar{z})
= - {\partial \over 
\partial \bar{z}} \left( { \partial z_m \over \partial m^a} (z,\bar{z}) 
\right) \ . \efr
In string theory it is customary to introduce holomorphic coordinates on 
moduli space, i.e. choose $3g-3+N$ complex parameters $m^I$, with 
complex conjugates $m^{*I}$, such that 
$$ {\partial z_m (\xi) \over \partial m^{*I}} = 0 \ . \efr
Then we obtain the correct integration measure on moduli space by 
inserting into the path integral over the right-moving
reparametrization ghosts $b=b_{zz}$ and $c=c^z$ the $3g-3+N$ 
conformally invariant factors
$$ (\eta_{m^I} \vert b) = \int {{\rm d}^2z \over \pi} (\eta_{m^I} 
)_{\bar{z}}^{\ z} (z,\bar{z}) b_{zz}(z) \ , \efr
with similar factors for the left-movers. 

So far we have only considered local coordinates on moduli space.
To have a {\it global} description it is convenient to consider instead
Teichm\"{u}ller space, which is the space of complex structures modulo
diffeomorfisms {\it continuously connected to the identity}. Unlike
moduli space, Teichm\"{u}ller space is actually a complex manifold (in
the strict mathematical sense); moreover, it is simply connected and can be
covered by a single global set of coordinates, which may even be
chosen to be everywhere holomorphic~[\Ref{Nag}]. Thus, the points of
Teichm\"{u}ller space are in one-to-one correspondence with the points
of the open domain ${\cal O} \in {\Bbb C}^{3g-3+N}$ where this coordinate is
defined. By definition each point in Teichm\"{u}ller space corresponds
to a set of complex structures related to each other (like in
eq.~\diffequiv ) by diffeomorfisms continuously connected to the identity.
Two different points $m$ and $m'$ in Teichm\"{u}ller space
will correspond to the same point in moduli space whenever the complex
structures pertaining to one point are related to those
pertaining to the other point by means of a diffeomorfism that is {\it not}
continuously connected to the identity. The group of such
diffeomorfisms (modulo the group of diffeomorfisms continuously
connected to the identity) is called the group of modular
transformations. It acts on the points of Teichm\"{u}ller space and
this allows us to identify moduli space with any subset ${\cal D}$ of
${\cal O}$ which is a fundamental domain of the group of modular
transformations.~\note{The group of modular transformations also acts
on the homology cycles, and it is often very convenient to describe a
modular transformation by its action on a canonical homology
basis. However, this is only a partial description, because there
exist modular transformations (the so-called Torelli subgroup) which
act trivially on the homology basis.}
\section{Mirror image Riemann surfaces.}
By taking the complex conjugate of the Beltrami equation \Beltrami \ we 
see that if $z(\xi)$ is a holomorphic coordinate pertaining to the 
complex structure $J$, then 
$$ \tilde{z} (\xi) \equiv (z(\xi))^* \nfr{tildez}
is a holomorphic coordinate pertaining to the complex structure 
$\tilde{J}$ defined by
$$ \tilde{J}_r^{\ s} (\xi) \equiv - \left( J_r^{\ s} (\xi) \right)^* =
- J_r^{\ s} (\xi) \ , \nfr{tildej}
where the last equality sign follows from the fact that the $J_r^{\ s}$ 
are real-valued functions. 

We call the complex structure $\tilde{J}$ the {\it mirror image} of $J$, 
and the Riemann surface $\tilde{M} = ({\cal M},\tilde{J})$ 
the {\it mirror image} of the Riemann surface $M=({\cal M},J)$.
By construction all holomorphic coordinates on $M$ are
anti-holomorphic functions of the holomorphic coordinates on $\tilde{M}$.
The two Riemann surfaces describe different points in moduli space 
unless $J \congphi \tilde{J}$, 
i.e. unless the two complex structures are 
related by a diffeomorfism $\Phi$, as in eq.~\diffequiv . Obviously,
$M \rightarrow \tilde{M}$ is a one-to-one map of moduli space
onto itself, since performed twice it is just the identity map.

By multiplying both sides of eq.~\diffequiv \ with minus one we see that
if the complex structures $I$ and $J$ are related by the diffeomorfism 
$\Phi$, so are the complex structures $\tilde{I}$ and $\tilde{J}$, and 
vice versa. In short, 
$$I \congphi J \ \Leftrightarrow 
\ \tilde{I} \congphi \tilde{J} \ . \nfr{theorem}
This means that, given a local coordinate $m \rightarrow J_m$ on 
moduli space, as described in the previous subsection, 
we can define another equally good local coordinate by the map $m 
\rightarrow \tilde{J}_m$. 

When we consider the dependence of the complex structure on a 
holomorphic set of moduli, we have
$$ J_r^{\ s} (\xi) = J_r^{\ s} (m^I,m^{*I};\xi) \ . \nfr{mapone}
Since $J_r^{\ s}$ is real-valued it has to depend on both $m^I$ and 
$m^{*I}$. But then, since $\tilde{J}$ is defined from $J$ by means of 
complex conjugation, as in eq.~\tildej ,
it is a function of $m^{*I}$ and $m^I$, rather than
$m^I$ and $m^{*I}$:
$$ \tilde{J}_r^{\ s} (\xi) = - (J_r^{\ s})^* (m^{*I},m^I;\xi) \equiv 
\tilde{J}_r^{\ s} (m^{*I},m^I;\xi) \ .
\nfr{maptwo}
In a convenient shorthand notation, the map \mapone \
is represented by $m \rightarrow J_m$ and the map \maptwo \ 
by $m^* \rightarrow \tilde{J}_{m^*}$. 

In summary, the situation is the following: Given a local holomorphic
coordinate system ${\cal D} 
\ni m \rightarrow J_m$, then the 
map $J \rightarrow \tilde{J}$ relating a Riemann surface to its mirror 
image automatically defines a new local coordinate system ${\cal D}^* \ni m^* 
\rightarrow \tilde{J}_{m^*}$. If the Riemann surface $M$ is described 
by the point $m^I \in {\cal D}$ 
in the first set of coordinates, then the mirror image $\tilde{M}$ is 
described by the point $m^{*I} \in {\cal D}^*$ in the second set of 
coordinates.

By taking the complex conjugate of eq.~\beltramidif \ we find
$$\eqalignno{ & 
\left( \left( \eta_{m^I} \right)_r^{\ s} (m^I,m^{*I};\xi) \right)^*
\ = \ \left( {-1 \over 2i} {\partial \over \partial m^{I}} 
J_r^{\ s} (m^{I},m^{*I};\xi) \right)^* & \numali \cr
&\qquad\qquad\quad = \ {-1 \over 2i} {\partial \over \partial m^{*I}}
\tilde{J}_r^{\ s} (m^{*I},m^I;\xi) \ = \ \left(
\eta_{m^{*I}} \right)_r^{\ s} (m^{*I},m^I;\xi) \ , \cr } $$
which expresses the fact that the complex conjugate of the Beltrami 
differential pertaining to $m^I$ in the coordinate system $\{ m^I \}$ is
the Beltrami differential pertaining to $m^{*I}$ in the coordinate 
system $\{ m^{*I} \}$.

So, if we imagine a family of local coordinates $\{ m^I_1 \}$, $\{ m^I_2 \},
\ldots$, which covers exactly one fundamental domain of the modular
group inside Teichm\"{u}ller space, then $\{ m^{*I}_1 \}, \{ m^{*I}_2
\}, \ldots$ is another family of local coordinates
which will also cover exactly one fundamental domain. This follows
from the fact that the map $M
\rightarrow \tilde{M}$ relating a Riemann surface to its mirror image
is a one-to-one map of moduli space onto itself.
\references
\beginref
\Rref{Verlinde}{E.~Verlinde and H.~Verlinde, Phys.Lett. {\bf B192}
(1987) 95.}
\Rref{PDV1}{P.~Di Vecchia, M.L.~Frau, K.~Hornfeck, A.~Lerda, F.~Pezzella
and S.~Sciuto, Nucl.Phys. {\bf B322} (1989) 317.}
\Rref{PDV2}{P.~Di Vecchia, invited talk at the Workshop on
String quantum Gravity and Physics at the Planck scale, Erice,
June 1992, Int.Journ.Mod.Phys. {\bf A}.}
\Rref{Anto}{I.~Antoniadis, C.~Bachas, C.~Kounnas and P.~Windey,
Phys.Lett. {\bf 171B} (1986) 51; \newline
I.~Antoniadis, C.~Bachas and C.~Kounnas, Nucl.Phys. {\bf B289} 87;
\newline I.~Antoniadis and C.~Bachas, Nucl.Phys. {\bf B298} (1988) 586.}
\Rref{KLT}{H.~Kawai, D.C.~Lewellen and S.-H.H.~Tye, Nucl.Phys.
{\bf B288} (1987) 1;\newline
H.~Kawai, D.C.~Lewellen, J.A.~Schwartz and S.-H.H.~Tye, Nucl.Phys.
{\bf B299} (1988) 431.}
\Rref{Sonoda}{H.~Sonoda, Nucl.Phys. {\bf B326} (1989) 135.}
\Rref{Amano}{K.~Amano and A.~Tsuchiya, Phys.Rev. {\bf D39} (1989) 565;
\newline
K.~Amano, Nucl.Phys. {\bf B328} (1989) 510.}
\Rref{Mitchell}{D.~Mitchell, N.~Turok, R.~Wilkinson and P.~Jetzer,
Nucl.Phys. {\bf B315} (1989) 1; \newline
D.~Mitchell, B.~Sundborg and N.~Turok, Nucl.Phys. {\bf B335} (1990) 621.}
\Rref{Weisberger}{J.L.~Montag and W.I.~Weisberger, Nucl.Phys. {\bf B363}
(1991) 527.}
\Rref{Berera}{A.~Berera, Nucl.Phys. {\bf B411} (1994) 157.}
\Rref{Hoker}{K.~Aoki, E.~D'Hoker and D.H.~Phong, Nucl.Phys. {\bf B342}
(1990) 149;\newline
E.~D'Hoker and D.H.~Phong, Phys.Rev.Lett. {\bf 70} (1993) 3692; \newline 
Theor.Math.Phys. {\bf 98} (1994) 306 \hbox{[hep-th/9404128]};\newline 
Nucl.Phys. {\bf B440} (1995) 24 \hbox{[hep-th/9410152]}.}
\Rref{Bluhm}{R.~Bluhm, L.~Dolan and P.~Goddard, Nucl.Phys. {\bf B309}
(1988) 330.}
\Rref{mink}{A.~Pasquinucci and K.~Roland, Phys.Lett. {\bf B351} (1995) 131 
\hbox{[hep-th/9503040]}.}
\Rref{normaliz}{A.~Pasquinucci and K.~Roland, Nucl.Phys. {\bf B457} (1995) 
27 \hbox{[hep-th/9508135]}.}
\Rref{SW}{R.F.~Streater and A.S.~Wightman, ``{\sl PCT, Spin and Statistics,
and all that\/},'' Benjamin Inc., New York, 1964.}
\Rref{Luders}{G.~L\"uders, Ann.Phys. {\bf 2} (1957) 1.}
\Rref{Pauli}{W.~Pauli, ``{\sl Niels Bohr and the Development of Physics\/}", 
McGraw-Hill, New York, 1955;
NuovoCimento {\bf 6} (1957) 204.}
\Rref{uscpt}{A. Pasquinucci and K. Roland, Nucl.Phys. {\bf B473}
(1996) 31  \hbox{[hep-th/9602026]}.}
\Rref{others}{J.~Schwinger, Phys.Rev.{\bf 82} (1951) 914; \newline
J.S.~Bell, Proc.Roy.Soc. (London) {\bf A231} (1955) 479; \newline
G.~L\"{u}ders and B.~Zumino, Phys.Rev. {\bf 106} (1957) 345; \newline
R.~Jost, ``{\it The General Theory of Quantized Fields},'' AMS, Providence, 
1965.}
\Rref{BPZ}{A.A.~Belavin, A.M.~Polyakov and A.B.~Zamolodchikov, 
Nucl.Phys. {\bf B241} (1984) 333.}
\Rref{MooreSeiberg}{G.~Moore and N.~Seiberg, Nucl.Phys. {\bf B313} (1989) 16;
Comm.Math.Phys. {\bf 123} (1989) 177.}
\Rref{Sonodatwo}{H.~Sonoda, Nucl.Phys. {\bf B311} (1988/89) 401; 
Nucl.Phys. {\bf B311} (1988/89) 417.}
\Rref{Bers}{L.~Bers, Bull. London Math. Soc. {\bf 4} (1972) 257.}
\Rref{Nag}{S.~Nag, ``{\it The Complex Analytic Theory of
Teichm\"{u}ller Spaces}'', Wiley, 1988.}

\endref
\ciao
